\documentclass[amsmath,amssymb,pra,superscriptaddress]{revtex4-1}
\usepackage{graphicx}
\usepackage{dcolumn}
\usepackage{bm}
\usepackage{subfigure}
\usepackage{booktabs}
\usepackage{color}
\newcounter{one}
\newcommand{\pddt}{\frac{ \partial }{ \partial \tau} }
\newcommand{\ddtau}{\frac{ d }{ d\tau}}
\newcommand{\bra}[1]{\langle #1 |}
\newcommand{\ket}[1]{| #1 \rangle}
\newcommand{\braket}[1]{\langle #1 \rangle}
\newcommand{\brabra}[1]{\langle\!\langle #1 |}
\newcommand{\ketket}[1]{| #1 \rangle\!\rangle}
\newcommand{\brabraketket}[2]{\langle\!\langle #1 | #2 \rangle\!\rangle}
\newcommand{\hathat}[1]{ \hat{\hat{ #1 }} }

\newcommand{\affA}{
Department of Physics, The University of Tokyo, Komaba, Meguro, Tokyo 153-8505 
}
\newcommand{\affB}{
Institute of Industrial Science, The University of Tokyo, Komaba, Meguro, Tokyo 153-8505
}

\begin{document}

\title{\textbf{A test of ``fluctuation theorem" in non-Markovian open quantum systems}}

\author{Tatsuro Kawamoto}
\affiliation{\affA}
\author{Naomichi Hatano}
\affiliation{\affB}
\date{\today}

\begin{abstract}
We study fluctuation theorems for open quantum systems with a non-Markovian heat bath using the approach of quantum master equations and examine the physical quantities that appear in those fluctuation theorems.
The approach of Markovian quantum master equations to the fluctuation theorems was developed by Esposito and Mukamel
[Phys.\,Rev.\,E {\bf73}, 046129 (2006)].
We show that their discussion can be formally generalized to the case of a non-Markovian heat bath when the local system is linearly connected to a Gaussian heat bath with the spectrum distribution of the Drude form.
We found by numerically simulating the spin-boson model in non-Markovian regime that the ``detailed balance" condition is well satisfied except in a strongly non-equilibrium transient situation, and hence our generalization of the definition of the ``entropy production" is almost always legitimate. 
Therefore, our generalization of the fluctuation theorem seems meaningful in wide regions.


\end{abstract}
 
\maketitle

\section{Introduction}
The research of non-equilibrium statistical physics has been energetically developed for the last decade in the context of fluctuation theorems and the Jarzynski equality
%The fluctuation theorem for a quantum system was first discussed for essentially isolated systems
\cite{2000cond.mat..7360K,2000cond.mat..9244T,PhysRevE.75.050102,1751-8121-40-26-F08,PhysRevLett.100.230404,PhysRevE.72.027102,2003cond.mat..8337M,JPSJ.69.2367}.
In particular, the case of open quantum systems,
where a reservoir with an infinite number of degrees of freedom is attached to a local quantum system with a finite number of degrees of freedom, 
is of great interest presently \cite{PhysRevE.81.011129,Talkner:2009fr,PhysRevLett.102.210401,PhysRevE.75.050102,RevModPhys.81.1665,1742-5468-2008-10-P10023,PhysRevE.73.046129,2011arXiv1103.4775D,2010PhRvL.105n0601C}.
The most common approach to discussing the fluctuation theorems for an open quantum system is by the twice energy measurement on the local system together with the reservoir.
There, one often assumes that the coupling between the local system and the reservoir is weak.
This approach, since the measurements are done on both the local system and the reservoir, 
can be viewed as manipulating the total isolated system.

In contrast, we here consider whether there exist any fluctuation theorems for an open quantum system that are  expressed solely in terms of quantities of its local system.
Fluctuation theorems of such kind were indeed proposed by Esposito and Mukamel \cite{PhysRevE.73.046129}
under the context of Markovian quantum master equations.
Their approach is a quantum analog to the fluctuation theorems of classical stochastic processes discussed by Crooks \cite{springerlink:10.1023/A:1023208217925,PhysRevE.60.2721} and Seifert \cite{PhysRevLett.95.040602}.
Note, however, that the entropy production that appears in their formalism is not a thermodynamic entropy production;
it is an information entropy production.
We will use the term entropy production in this sense in the following.

In the present paper, 
we will first show that we can formally generalize the discussion by Esposito and Mukamel to the case of the dynamics in a \textit{non}-Markovian heat bath.
To this end, we employ a method called hierarchy equations of motion, which describes the reduced dynamics in a non-Markovian heat bath.
The generalization to the non-Markovian dynamics in classical systems was already done in Refs.~\cite{1742-5468-2005-09-P09013,1742-5468-2007-10-P10010, 1742-5468-2007-09-L09002}.

When we discuss the fluctuation theorems, defining the work, the heat, and the entropy production is also an  essential problem \cite{2003cond.mat..8337M,PhysRevE.71.066102,1367-2630-12-1-013013,2010LNP...784.....G,2011arXiv1103.4775D}.
As was done by Esposito and Mukamel \cite{PhysRevE.73.046129}, we can proceed with formal definitions based on quantum analogs to classical stochastic processes as working assumptions.
However, it is necessary to examine whether the definitions are indeed appropriate quantum-mechanically.
We will examine the validity of the ``entropy production," the ``entropy flow," and the ``heat" that appear in the fluctuation theorems under our generalization by analyzing the spin-boson model, numerically in the non-Markovian regions and analytically in the limit where the Born-Markov approximation and the rotating-wave approximation are applicable. 
We will find that the ``entropy production" indeed vanishes in equilibrium, 
and thus we can regard that our fluctuation theorem is legitimate even in non-Markovian regions except for strongly non-equilibrium transient states. 

As another point, if the ``entropy flow" is related to the ``heat" by the ``microscopic reversibility," then we can rewrite our fluctuation theorem into the Crooks-type when the initial and the final states are the canonical states. 
We will find, however, that the ``microscopic reversibility" is broken in wide parameter regions although it does hold in the limit where the Born-Markov approximation and the rotating-wave approximation are applicable in the driving protocols that we will investigate in the present paper.
Furthermore, we will find that the ``detailed balance" is almost satisfied in some parameter regions.

The paper is organized as follows.
In Sec.\,\ref{main1}, we will derive the fluctuation theorems for a quantum master equation in a non-Markovian heat bath.
In this Section, we will assume that the heat bath is composed of the ensemble of harmonic oscillators that are linearly connected to the local system, but the Hamiltonian of the local system can be arbitrary.
In Sec.\,\ref{main2}, we will show our numerical results for the spin-boson model;
we will demonstrate the time evolution of the deviation from the ``detailed balance." 
In equilibrium states, the ``detailed balance" condition means that the ``entropy production" is zero. 
We will also analyze the case where the local system is weakly driven by a Zeeman magnetic field periodically.
The ``entropy flow" and the ``entropy production" are well-defined in this case, although they might be ill-defined when we drive the local system strongly or in different schedules. 
We will also examine whether any other useful relation holds; we will find that, although the ``microscopic reversibility" which we will discuss in detail below, does not hold in general, there exist a case where the equation of the ``detailed balance" instead is almost satisfied.

\section{Fluctuation theorems for a quantum master equation in a non-Markovian heat bath}\label{main1}
We consider a local system which is linearly connected to a Gaussian heat bath.
The Hamiltonian of the total system reads
\begin{align}
&\hat{H}=\hat{H}_{\mathrm{S}}+\hat{H}_{\mathrm{B}}+\hat{H}_{\mathrm{int}}, \label{totalHamiltonian}\\
&\hat{H}_{\mathrm{B}}(\hat{b}_{\alpha}, \hat{b}_{\alpha}^{\dagger})
=\sum_{\alpha} \hbar \omega_{\alpha} \hat{b}_{\alpha}^{\dagger} \, \hat{b}_{\alpha}, \label{HB}\\
&\hat{H}_{\mathrm{int}}
=\hat{V} \sum_{\alpha} \widetilde{c}_{\alpha} \hat{x}_{\alpha} 
=: \hat{V} \sum_{\alpha} c_{\alpha} (\hat{b}_{\alpha} + \hat{b}_{\alpha}^{\dagger}).\label{Hint}
\end{align}
The Hamiltonian of the local system $\hat{H}_{\mathrm{S}}$ is arbitrary.
The heat-bath Hamiltonian $\hat{H}_{\mathrm{B}}(\hat{b}_{\alpha}, \hat{b}_{\alpha}^{\dagger})$ consists of the harmonic oscillators of an infinite number of modes $\omega_{\alpha}$, where
$\hat{b}_{\alpha}^{\dagger}$ and $\hat{b}_{\alpha}$ are the creation and annihilation operators of each mode.
We omitted the ground-state energy of the harmonic oscillators.
The interaction between the system and the heat bath $\hat{H}_{\mathrm{int}}$ is linear in the position of the harmonic oscillator of each mode 
$\hat{x}_{\alpha}=\sqrt{\hbar/2m_{\alpha}\omega_{\alpha}} \,(\hat{b}_{\alpha}^{\dagger}+\hat{b}_{\alpha})$
with the coupling strength $\widetilde{c}_{\alpha}=c_{\alpha} \sqrt{2m_{\alpha}\omega_{\alpha}/\hbar}$.
The operator $\hat{V}$ is an arbitrary one of the local system which depends on time in general. 
%The total Hamiltonian can be recast as
%\begin{align}
%&\hat{H}= \frac{\hbar \omega_{0}}{2} ( \hat{\psi}^{\dagger} \hat{\psi} - \hat{\psi} \hat{\psi}^{\dagger} )
%+ \sum_{\alpha} \left[ \frac{\hat{p}_{\alpha}^{2}}{2m_{\alpha}} 
%+\frac{1}{2} m_{\alpha} \omega_{\alpha}^{2} \left( \hat{x} - \frac{c_{\alpha} V(\hat{\psi}, \hat{\psi}^{\dagger})}{m_{\alpha} \omega_{\alpha}^{2}} \right)^{2} \right].\label{totalHamiltonian2}
%\end{align}
We approximate that the spectral distribution of the bath mode 
$J(\omega)=\sum_{\alpha} ( c_{\alpha}^{2}\hbar / 2 m_{\alpha}\omega_{\alpha} ) \delta(\omega-\omega_{\alpha})$ 
is of the Drude form:
\begin{align}
J(\omega) 
%=\sum_{\alpha} \frac{c_{\alpha}^{2}\hbar}{2 m_{\alpha}\omega_{\alpha}}
% \delta(\omega-\omega_{\alpha})
= \frac{\hbar^{2} \zeta}{\pi \omega_{0}} \frac{\gamma^{2} \omega}{\omega^{2} + \gamma^{2}},
\label{DrudeForm}
\end{align}
which is the Ohmic distribution with the Lorentzian cutoff $\gamma$ and a coefficient $\zeta$.
The cutoff $\gamma$ gives the decay rate of the canonical time correlation function of the heat bath, 
whereas $\zeta$ is the coefficient related to the system-bath coupling strength $c_{\alpha}$. 
For the region where the Born-Markov approximation is appropriate, 
i.e.\,, 
\begin{align}
&\min \left[ \gamma, \frac{2 \pi}{\beta} \right] \gg \zeta, \label{BornMarkovCondition}
\end{align}
we refer to it as a Markovian heat bath; 
otherwise we refer to it as a \textit{non}-Markovian heat bath.

Esposito and Mukamel \cite{PhysRevE.73.046129} started their discussion with a Markovian quantum master equation of the form
\begin{equation}
\ddtau \ketket{ \hat{\rho}(\tau) } = \hat{\hat{K}}(\tau) \ketket{ \hat{\rho}(\tau) } \label{EM1},
\end{equation}
where 
$\ketket{ \hat{\rho}(\tau) }$ is the reduced density matrix of the local system and 
they used the following notation:
\begin{align}
\ketket{a,b} &\equiv \ket{a}\bra{b}, \\
\brabraketket{a,b}{c,d} &\equiv \langle a \ket{c} \langle d \ket{b}. \label{EM2}
\end{align}
In (\ref{EM1}), $\hat{\hat{K}}$ is a dynamical semi-group acting on the reduced Liouville space of the local system.
(We use the single-hatted letter as an operator acting on the Hilbert space and the double-hatted letter as an operator acting on the Liouville space.)
The evolution of the reduced density matrix does not directly lead to the fluctuation theorem.
In order to investigate the fluctuation theorem analogously to classical stochastic processes, they translated the Markovian quantum master equation in a form of a \textit{classical} master equation using a time-dependent basis. 
They then constructed a ``quantum trajectory" for the dynamics and discussed the forward and backward probabilities of the trajectory.
The time-dependent basis is a basis which diagonalizes the reduced density matrix at each time,
and hence the reduced density matrix is represented as
\begin{align}
\langle m^{\prime}_{\tau}| \hat{\rho}(\tau) |m_{\tau} \rangle 
&= \brabraketket{ m^{\prime}_{\tau} m_{\tau} }{ \hat{\rho}(\tau) } = P_{\tau}(m) \delta_{m^{\prime} m}, \nonumber \\
\ketket{ \hat{\rho}(\tau) }&= \sum_{m} \ket{m_{\tau}}P_{\tau}(m)\bra{m_{\tau}}, \label{EM3}
\end{align}
where we suppressed the subscript $\tau$ for $m_{\tau}$ and $m^{\prime}_{\tau}$ on the right-hand side. 
Then we can regard the basis $\{ \ket{m_{\tau}} \}$ as a set of states with probability $\{P_{\tau}(m)\}$ at time $\tau$.
By connecting these states, Esposito and Mukamel were able to construct a quantum trajectory (Fig.\,1 in Ref.~\cite{PhysRevE.73.046129}) of the local system.  
Under this basis, they \cite{PhysRevE.73.046129} had the classical master equation representation of the quantum master equation (\ref{EM1}):
\begin{align}
\frac{d P_{\tau}(m)}{d\tau} &= \sum_{m^{\prime} (\ne m)} \bigl(
W_{\tau}(m,m^{\prime}) P_{\tau}(m^{\prime}) - W_{\tau}(m^{\prime},m) P_{\tau}(m)
\bigr), \label{EM4}
\end{align}
where
\begin{align}
W_{\tau}(m,m^{\prime}) &\equiv \brabra{ m_{\tau}m_{\tau} } \hat{\hat{K}}(\tau) \ketket{ m^{\prime}_{\tau}m^{\prime}_{\tau} }. \label{EM5}
\end{align}

We will show that the discussion above by Esposito and Mukamel \cite{PhysRevE.73.046129} can be formally generalized to the dynamics in the non-Markovian heat bath.
In their discussion, it was essential that the equation of motion is expressed in a time-local form.
In order to keep the time-local form for the non-Markovian heat bath,
we employ a set of master equations called \textit{hierarchy equations of motion} 
\cite{JPSJ.58.101,JPSJ.74.3131,JPSJ.75.082001,PhysRevA.43.4131}.
In the formalism of the hierarchy equations of motion,
a state of the system is expressed with a set of infinite matrices in an extended space instead of a reduced density matrix.
Thereby we can obtain a time-local equation of motion by taking account of the time correlation of the heat bath in the form of correlations among the matrices.

\subsection{Hierarchy equations of motion}
In the formalism of the hierarchy equations of motion \cite{JPSJ.58.101,JPSJ.74.3131,JPSJ.75.082001,PhysRevA.43.4131},
a state of the local system is expressed as a set of an infinite number of matrices as follows:
\begin{align}
%&\left\{ \hat{\rho}^{(n)}_{j_{1},\dots, j_{k},\dots}(\tau) \right\} 
%= \hat{\rho}^{(0)}_{0,0,\dots}(\tau),\, \hat{\rho}^{(1)}_{0,0,\dots}(\tau),\, \hat{\rho}^{(1)}_{1,0,\dots}(\tau), \cdots,\\
&\hat{\rho}^{(0)}_{0,0,\dots}(\tau) \otimes \hat{\rho}^{(1)}_{0,0\dots}(\tau) \otimes \hat{\rho}^{(1)}_{1,0\dots}(\tau) \otimes \cdots \otimes \hat{\rho}^{(n)}_{j_{1},\dots, j_{k},\dots}(\tau) \otimes \cdots \nonumber\\
&=:\ketket{\hat{\rho}^{(0)}_{0,0,\dots}(\tau),\, \hat{\rho}^{(1)}_{0,0\dots}(\tau),\, \hat{\rho}^{(1)}_{1,0\dots}(\tau), \cdots}
=: \ketket{ \hat{\rho}(\tau) ; \{ \hat{\sigma}_{l}(\tau) \} }, \nonumber\\
&\hat{\rho}^{(0)}_{0,0,\dots}(\tau) \equiv \hat{\rho}(\tau), \hspace{5mm}
 \hat{\rho}^{(1)}_{0,0,\dots}(\tau) \otimes \hat{\rho}^{(1)}_{1,0,\dots}(\tau) \cdots 
=: \{ \hat{\sigma}_{l}(\tau) \},
\end{align}  
where $\hat{\rho}^{(0)}_{0,0\dots}(\tau)$ is the reduced density matrix $\hat{\rho}(\tau)$ in the usual sense,
while $\{ \hat{\sigma}_{l}(\tau) \}$ are the set of auxiliary matrices which possess the information of non-Markovian effects.
These matrices follow from the equations of motion (\ref{hierarchyeqn0})--(\ref{hierarchyeqn}) in Appendix \ref{HierarchyReview}. 
The auxiliary matrices $\{ \hat{\sigma}_{l}(\tau) \}$ are introduced for computational purposes only and possess no physical meaning themselves.
The superscript $n$ indicates the $n$th correction with respect to the decay rate of the heat bath $\gamma$
and the subscript $\{j_{k}|k=1,2,\dots\}$ indicates the corrections with respect to the temperature, i.e.\,the Matsubara frequencies $\nu_{k}=2\pi k/\beta\hbar$.
%Both of these corrections are neglected in the Markovian approximation.
We define its inner product as
\begin{align}
&\brabraketket{ \hat{\rho}(\tau) ; \{ \hat{\sigma}_{l}(\tau) \} }
{\hat{\rho}^{\prime}(\tau) ; \{ \hat{\sigma}^{\prime}_{l}(\tau) \} }\nonumber\\
&\equiv
\mathrm{Tr}\, \left( \hat{\rho}^{\dagger}(\tau) \hat{\rho}^{\prime}(\tau)\right)
+\sum_{l=1}^{\infty} \, \mathrm{Tr}\, \left( \hat{\sigma}^{\dagger}_{l}(\tau)  \hat{\sigma}^{\prime}_{l}(\tau) \right).\label{innerproduct}
\end{align}
Then the hierarchy equations of motion (\ref{hierarchyeqn0})--(\ref{hierarchyeqn}) can be formally expressed as
\begin{align}
&\ddtau \ketket{ \hat{\rho}(\tau) ; \{ \hat{\sigma}_{l}(\tau) \} } = \hathat{L}_{ \mathrm{hier} }(\tau) \ketket{ \hat{\rho}(\tau) ; \{ \hat{\sigma}_{l}(\tau) \} }, \label{HQME1}
\end{align}
where $\hathat{L}_{ \mathrm{hier} }(\tau)$ is the generator that represents all the operations on the right-hand sides of Eqs.\,(\ref{hierarchyeqn0})--(\ref{hierarchyeqn}).
Note that Eq.~(\ref{HQME1}) is derived under the assumption that the local system is linearly coupled to the Gaussian heat bath (\ref{HB}) with the spectrum distribution of the Drude form (\ref{DrudeForm}).
This setting is often assumed when we derive a Markovian quantum master equation microscopically.

\subsection{Birth-death master equation for the hierarchy equations of motion}\label{BDMEforHierarchy}
Analogously to Ref.~\cite{PhysRevE.73.046129}, 
let $\{ \ket{m_{\tau}, m_{\tau}; \{ \hat{0} \}} \}$ be the states that satisfy
%a time-dependent basis that diagonalizes the reduced density matrix $\hat{\rho}(\tau)$
%and let $P_{\tau}(m)$ refer to its eigenvalue, i.e.\,
\begin{align}
\brabraketket{ m_{\tau}, m_{\tau}; \{ \hat{0} \} }{ \hat{\rho}(\tau) ; \{ \hat{\sigma}_{l}(\tau) \} }
&=\mathrm{Tr}\, \ket{m_{\tau}} \bra{m_{\tau}} \hat{\rho}(\tau) \ket{m_{\tau}} \bra{m_{\tau}} \nonumber\\
&= P_{\tau}(m), \label{td-basis}
\end{align}
where $\{ \hat{0} \} = \{\hat{0},\hat{0},\dots \}$ denotes an infinite set of zero matrices in the space of the auxiliary matrices.
Our purpose is to extract out the equation of motion for the probability $P_{\tau}(m)$. 
Taking the inner product with  $\brabra{m_{\tau}m_{\tau};\{ \hat{0} \}}$ in (\ref{HQME1}), we have
\begin{align}
&\brabra{m_{\tau}m_{\tau};\{ \hat{0} \}} \ddtau \ketket{ \hat{\rho}(\tau) ; \{ \hat{\sigma}_{l}(\tau) \} } \nonumber\\
&\hspace{5mm}= \brabra{m_{\tau}m_{\tau};\{ \hat{0} \}}
\hat{\hat{ \mathcal{L} }}_{ \mathrm{hier} }(\tau) \ketket{ \hat{\rho}(\tau) ; \{ \hat{\sigma}_{l}(\tau) \} }. \label{HQME2}
\end{align}
Next, we decompose $\ketket{ \hat{\rho}(\tau) ; \{ \hat{\sigma}_{l}(\tau) \} }$ into the following form:
\begin{align}
\ketket{ \hat{\rho}(\tau) ; \{ \hat{\sigma}_{l}(\tau) \} }
&=\sum_{m} \left[\left( \bra{m_{\tau}} \hat{\rho}(\tau) \ket{m_{\tau}} \right) \ket{m_{\tau}} \bra{m_{\tau}} \otimes \prod_{l=1}^{\infty} \hat{\sigma}_{l}(\tau) \right] \nonumber\\
&=\sum_{m}  \ketket{ m_{\tau} m_{\tau} ; \{ \hat{\sigma}_{l}(\tau) \} } P_{\tau}(m).\label{decomposition}
\end{align}
In the first equality of (\ref{decomposition}), we used
the completeness relation for the reduced density matrix $\hat{\rho}(\tau)$.
In the formalism of Ref.~\cite{PhysRevE.73.046129}, it reads
\begin{align}
\ketket{ \hat{\rho}(\tau) }
&= \sum_{m} \left( \ketket{m_{\tau},m_{\tau}}\brabra{m_{\tau},m_{\tau}} \right) \ketket{ \hat{\rho}(\tau) } \nonumber\\
&=\sum_{m} \ket{m_{\tau}} \bra{m_{\tau}} \hat{\rho}(\tau) \ket{m_{\tau}} \bra{m_{\tau}} .
\end{align}
Using (\ref{decomposition}) in the right-hand side of (\ref{HQME2}), we have
\begin{align}
%\ddtau &\brabraketket{ m_{\tau}m_{\tau}; \{\hat{0}\} }{ \hat{\rho}(\tau) ; \{ \hat{\sigma}_{l}(\tau) \} }\nonumber\\
&\brabra{m_{\tau}m_{\tau};\{ \hat{0} \}} \ddtau \ketket{ \hat{\rho}(\tau) ; \{ \hat{\sigma}_{l}(\tau) \} }\nonumber\\
&\hspace{5mm}= \sum_{m^{\prime}} \brabra{ m_{\tau}m_{\tau}; \{\hat{0}\} }
\hat{\hat{ \mathcal{L} }}_{ \mathrm{hier} }(\tau) \ketket{ m^{\prime}_{\tau}m^{\prime}_{\tau} ; \{ \hat{\sigma}_{l}(\tau) \} }
P_{\tau}(m^{\prime})
. \label{HQME3'}
\end{align}
The left-hand side of (\ref{HQME3'}) can be also recast as
\begin{align}
&\brabra{m_{\tau}m_{\tau};\{ \hat{0} \}} \ddtau \ketket{ \hat{\rho}(\tau) ; \{ \hat{\sigma}_{l}(\tau) \} } \nonumber\\
&\hspace{5mm}=\mathrm{Tr}\, \left( \ket{m_{\tau}}\bra{m_{\tau}} \ddtau \hat{\rho}(\tau)\right)
+\sum_{k=1}^{\infty} \, \mathrm{Tr}\, \left( \hat{0}\,  \ddtau \hat{\sigma}_{l}(\tau) \right) \nonumber\\
&\hspace{5mm}=\ddtau P_{\tau}(m) - \left( \ddtau \bra{m_{\tau}} m_{\tau}\rangle \right) P_{\tau}(m) \nonumber\\
&\hspace{5mm}=\ddtau P_{\tau}(m).
\end{align}
Then we have
\begin{align}
\frac{d P_{\tau}(m)}{d\tau}
&= \sum_{m^{\prime}} \brabra{ m_{\tau}m_{\tau}; \{\hat{0}\} }
\hat{\hat{ \mathcal{L} }}_{ \mathrm{hier} }(\tau) \ketket{ m^{\prime}_{\tau}m^{\prime}_{\tau} ; \{ \hat{\sigma}_{l}(\tau) \} }
P_{\tau}(m^{\prime})
. \label{HQME3}
\end{align}
Defining 
\begin{align}
W_{\tau}(m, m^{\prime}; \{ \hat{\sigma}_{l}(\tau) \})
\equiv \brabra{ m_{\tau}m_{\tau} ; \{\hat{0}\}}
\hat{\hat{ \mathcal{L} }}_{ \mathrm{hier} }(\tau) \ketket{ m^{\prime}_{\tau}m^{\prime}_{\tau} ; \{ \hat{\sigma}_{l}(\tau) \} },
\label{HQMErate}
\end{align}
we can rewrite (\ref{HQME3}) as
\begin{align}
\ddtau P_{\tau}(m)
&= \sum_{m^{\prime}} W_{\tau}(m, m^{\prime} ; \{ \hat{\sigma}_{l}(\tau) \} ) \,
P_{\tau}( m^{\prime} ). 
\end{align}
Conservation of the probability during the evolution leads to
\begin{align}
0=\ddtau \mathrm{Tr}\, \hat{\rho}(\tau)
&= \ddtau \sum_{m} P_{\tau}(m) \nonumber\\
 &= \sum_{m^{\prime}} \left( \sum_{m} W_{\tau}(m,m^{\prime} ; \{ \hat{\sigma}_{l}(\tau) \} )\right) P_{\tau}(m^{\prime}). \label{ProbConservation}
\end{align}
Then we have
\begin{align}
&\sum_{m} W_{\tau}(m,m^{\prime} ; \{ \hat{\sigma}_{l}(\tau) \})=0,
\end{align}
i.e.
\begin{align}
W_{\tau}(m^{\prime},m^{\prime} ; \{ \hat{\sigma}_{l}(\tau) \})
=-\sum_{m\ne m^{\prime}} W_{\tau}(m,m^{\prime} ; \{ \hat{\sigma}_{l}(\tau) \}).
\end{align}
We thereby obtain an equation of motion for the hierarchy equations of motion
in the form of a classical birth-death master equation:
\begin{align}
\ddtau P_{\tau}(m)
= \sum_{m^{\prime} (\ne m)} \biggl[
&W_{\tau}(m,m^{\prime} ; \{ \hat{\sigma}_{l}(\tau) \} ) P_{\tau}(m^{\prime}) \nonumber\\
& - W_{\tau}(m^{\prime},m ; \{ \hat{\sigma}_{l}(\tau) \}) P_{\tau}(m)
\biggr]. \label{HQME4}
\end{align}
This is a generalization of Eq.~(\ref{EM4}) to the case of a non-Markovian heat bath.
It is essential to consider an open system, since we can easily show that the ``transition rate" $W_{\tau}(m,m^{\prime}; \{ \hat{\sigma}_{l}(\tau) \})$ becomes zero for an isolated system; see Appendix \ref{AppendixIsolated}.
Indeed, the transition of the local system occurs thanks to the coupling to the heat bath.

Note, however, that the positivity of the ``transition rate" $W_{\tau}(b_{\tau},a_{\tau})$ in (\ref{HQMErate}) 
is not guaranteed.
For the birth-death master equation in classical stochastic processes, 
we construct the equation of motion based on physically given transition rates; the transition rate there is positive by definition. 
In contrast, Eq.~(\ref{HQME4}) is the equation that we derived from the Liouville-von Neumann equation of the total system;
Eq.~(\ref{HQME4}) does not necessarily have the meaning of the birth-death master equation in the sense of stochastic processes.

Algebraically, Eq.~(\ref{HQME4}) is correct in any parameter regions, except for the restriction that the local system is linearly coupled to the Gaussian heat bath with the spectrum distribution of the Drude form.
Our calculation starts from the exact Liouville-von Neumann dynamics of the total system and uses no approximations before arriving at (\ref{HQME4}).
Therefore, the simulations that we will show in Sec.~\ref{main2} are numerically exact.

\subsection{Forward and Backward Trajectories}\label{FBtrajectories}
\begin{figure}[t]
\begin{center}
\includegraphics[width=100mm,clip]{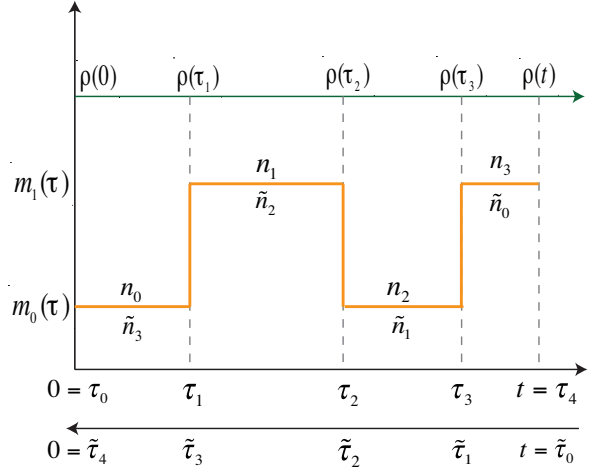}
\caption{
(color online) 
A quantum trajectory $n(\tau)$ for $N=3$ in the case of a two-level system.
Precisely speaking, the state $m_{0}$ and $m_{1}$ are time-dependent, and hence the axis should be dynamically modified, but it is suppressed here for simplicity.
The labels with tilde represent the backward process.}
\label{quantum_trajectory}
\end{center}
\end{figure}

Equation (\ref{HQME4}) naturally implies a quantum trajectory analogous to the trajectory in classical stochastic processes.
We can regard the basis $\{ \ket{m_{\tau}} \}$ as a set of states with probability $\{P_{\tau}(m)\}$ 
and the local system hops from a state to another with the ``transition rate" $W_{\tau}(m^{\prime},m ; \{ \hat{\sigma}_{l}(\tau) \})$
at time $\tau$.
%We label the elements of the basis $\ket{m_{\tau}}$ in the non-decreasing order of eigenvalues of $\hat{\rho}(\tau)$.
We can construct a forward quantum trajectory by
labeling the times $\tau_{j} \, (j=1,2,\cdots, N)$ when the transitions occur and
connecting these states (Fig.~\ref{quantum_trajectory}):
\begin{align}
n(\tau) = n_{0} \rightarrow n_{1}\rightarrow n_{2} \dots \rightarrow n_{N}, \label{forwardtrajectory}
\end{align}
where $n_{j}$ represents the state after the $j$th transition at time $\tau_{j}$.
Hence, $n(\tau)=n_{j}$ for $\tau_{j} \le \tau < \tau_{j+1}$ when $j<N$ and
$n(\tau)=n_{N}$ for $\tau_{N} \le \tau\le t$.
We set $\tau_{0}=0$ and $\tau_{N+1}=t$.

Similarly, we can construct a backward trajectory corresponding to the forward one.
The label of time for the backward trajectory is related to the forward one as $\widetilde{\tau}_{j} \leftrightarrow \tau_{N-j+1}$
and the value (duration) of the time is related as $\widetilde{\tau}_{j} = t - \tau_{N-j+1}$.
  The label of the state is related as $\widetilde{n}_{j-1}=n_{N-j+1}$,  so that the backward trajectory corresponding to $n(\tau)$ is (Fig.~\ref{quantum_trajectory})
\begin{align}
\widetilde{n}(\widetilde{\tau})&= \widetilde{n}_{0} \rightarrow \widetilde{n}_{1}\rightarrow \widetilde{n}_{2} \dots \rightarrow \widetilde{n}_{N} \nonumber\\
&=n_{N} \rightarrow n_{N-1}\rightarrow n_{N-2} \dots \rightarrow n_{0},
\end{align}
where we set $\tilde{\tau}_{0}=t$ and $\tilde{\tau}_{N+1}=0$.

The probabilities for the forward and backward quantum trajectories are obtained just by replacing 
$W_{\tau_{j}}(n_{j}, n_{j-1})$ with $W_{\tau_{j}}(n_{j}, n_{j-1} ; \{ \hat{\sigma}_{l}(\tau_{j}) \})$ 
in Eqs.~(62) and (66) of Ref.~\cite{PhysRevE.73.046129}:
\begin{align}
&\mu_{\mathrm{F}}[n_{(\tau)}] 
   =P_{0}(n_{0}) \nonumber\\
   &\hspace{5mm}\times
   \Biggl[ \prod_{i=1}^{N}
   \mathrm{exp} \left( -\sum_{m} \int_{\tau_{i-1}}^{\tau_{i}} d\tau^{\prime} \, W_{\tau^{\prime}}(m, n_{i-1} ; \{ \hat{\sigma}_{l}(\tau^{\prime}) \} ) \right) \nonumber\\
   &\hspace{4cm} \times W_{\tau_{j}}(n_{j}, n_{j-1} ; \{ \hat{\sigma}_{l}(\tau_{j}) \}) \Biggr] \nonumber\\
 &\hspace{5mm} \times \mathrm{exp} \left( -\sum_{m} \int_{\tau_{N}}^{t} d\tau^{\prime} \, W_{\tau^{\prime}}(m, n_{N} ; \{ \hat{\sigma}_{l}(\tau^{\prime}) \})  \right), \label{forwardProb}\\
&\mu_{\mathrm{B}}[\widetilde{n}_{(\widetilde{\tau})}] 
= \widetilde{P}_{0}(\widetilde{n}_{0}) \nonumber\\
&\hspace{5mm}\times 
   \Biggl[ \prod_{i=1}^{N}
   \mathrm{exp} \left( -\sum_{\widetilde{m}} \int_{\widetilde{\tau}_{i-1}}^{\widetilde{\tau}_{i}} d\tau^{\prime} \, 
   \widetilde{W}_{\tau^{\prime}}(\widetilde{m}, \widetilde{n}_{i-1}; \{ \widetilde{\hat{\sigma}_{l}}(\tau^{\prime}) \} )  \right) \nonumber\\   
   &\hspace{4cm} \times \widetilde{W}_{\widetilde{\tau}_{j}}(\widetilde{n}_{j}, \widetilde{n}_{j-1}; \{ \widetilde{\hat{\sigma}_{l}}(\widetilde{\tau}_{j}) \} ) \Biggr] \nonumber \\
 &\hspace{5mm} \times \mathrm{exp} \left( -\sum_{\widetilde{m}} \int_{\widetilde{\tau}_{N}}^{0} d\tau^{\prime} \, 
 \widetilde{W}_{\tau^{\prime}}(\widetilde{m}, \widetilde{n}_{N} ; \{ \widetilde{\hat{\sigma}_{l}}(\tau^{\prime}) \} )  \right) \label{backwardProb}.
\end{align} 
Hereafter, we set the probability of the final state of the forward trajectory equal to the probability of the initial state of the backward trajectory, i.e.\,$\widetilde{P}_{0}(\widetilde{n}_{0}) = P_{t}(n_{N})$.

In the argument in Ref.~\cite{PhysRevE.73.046129}, the dynamics that gives the backward process was a hypothetical one; the real time-reversed dynamics does not satisfy the relation
$\widetilde{W}_{\widetilde{\tau}}(\tilde{m}, \widetilde{m}^{\prime}) = W_{t-\widetilde{\tau}}(\widetilde{m}, \widetilde{m}^{\prime})$
because the quantum master equation with the Markov approximation breaks the time reversal symmetry.
In contrast, the backward process that we consider in (\ref{backwardProb}) is truly the process of the time-reversed dynamics because the hierarchy equations of motion formally solve the total system and do not break the time reversal symmetry.
Then we exactly have
\begin{align}
\widetilde{W}_{\widetilde{\tau}}(\widetilde{m}, \widetilde{m}^{\prime}; \{ \widetilde{\hat{\sigma}_{l}}(\widetilde{\tau}) \}) 
= W_{t-\widetilde{\tau}}(\widetilde{m}, \widetilde{m}^{\prime}; \{ \hat{\sigma}_{l}(t-\widetilde{\tau}) \}). \label{transitionrule}
\end{align}
Note that this condition dictates that the auxiliary matrices $\{ \hat{\sigma}_{l}(\tau) \}$ also evolve backward because the  dynamics of the heat bath must be also reversed.
Therefore, precisely speaking, the backward process which satisfies (\ref{transitionrule}) also becomes a hypothetical one if we truncate the hierarchy, although it would be effectively a real time-reversed process if the conditions (\ref{deltaapproxK}) and (\ref{terminatorcondition}) in Appendix \ref{HierarchyReview} are satisfied.

\subsection{Fluctuation theorems}\label{FTmain}
By repeating the same argument as in Ref.~\cite{PhysRevE.73.046129}, 
we can show the fluctuation theorem for the dynamics in a non-Markovian heat bath.

We define the entropy change $\Delta s(t)$, the ``entropy flow" $\Delta s_\mathrm{e}(t)$, 
and the ``entropy production" $\Delta s_\mathrm{i}(t)$ along a single trajectory as
\begin{align}
\Delta s(t) &\equiv \int_{0}^{t} d\tau\,  \left(
- \left. \frac{ \dot{P}_{\tau}(n) }{ P_{\tau}(n) }\right|_{n(\tau)}
- \sum_{j=1}^{N} \delta(\tau - \tau_{j}) \ln \frac{P_{\tau}(n_{j})}{P_{\tau}(n_{j-1})} \right)\nonumber\\
&= \ln P_{0}(n_{0}) - \ln P_{t}(n_{N}),\label{NMentropy}\\
\Delta s_{\mathrm{e}}(t) &\equiv - \int_{0}^{t} d\tau\,
\sum_{j=1}^{N} \delta(\tau - \tau_{j}) \ln \frac{W_{\tau}(n_{j}, n_{j-1}; \{ \hat{\sigma}_{l}(\tau) \})}{W_{\tau}(n_{j-1}, n_{j}; \{ \hat{\sigma}_{l}(\tau) \})} \nonumber\\
&=- \sum_{j=1}^{N} \ln \frac{W_{\tau_{j}}(n_{j}, n_{j-1}; \{ \hat{\sigma}_{l}(\tau_{j}) \})}{W_{\tau_{j}}(n_{j-1}, n_{j}; \{ \hat{\sigma}_{l}(\tau_{j}) \})},\label{NMentropyflow}\\
\Delta s_{\mathrm{i}}(t) &\equiv \int_{0}^{t} d\tau\, 
\Biggl( - \left. \frac{ \dot{P}_{\tau}(n) }{ P_{\tau}(n) }\right|_{n(\tau)} \nonumber\\
&\hspace{5mm} - \sum_{j=1}^{N} \delta(\tau - \tau_{j})
\ln \frac{P_{\tau}(n_{j}) W_{\tau}(n_{j-1}, n_{j}; \{ \hat{\sigma}_{l}(\tau) \}) }{ P_{\tau}(n_{j-1}) W_{\tau}(n_{j}, n_{j-1}; \{ \hat{\sigma}_{l}(\tau) \}) }
\Biggr)\nonumber\\
&=\Delta s(t) - \Delta s_{\mathrm{e}}(t).\label{NMentropyproduction}
\end{align}
This type of definitions were first introduced by Schnakenberg \cite{RevModPhys.48.571} and used by Seifert \cite{PhysRevLett.95.040602} in order to discuss the fluctuation theorems in classical stochastic processes.
Esposito and Mukamel \cite{PhysRevE.73.046129} extended these definitions for the classical dynamics to the ones for the Markovian quantum dynamics.  
Our definitions (\ref{NMentropy})--(\ref{NMentropyproduction}) are their generalization.

We also define the logarithm of the ratio of the probabilities for a forward trajectory and a backward trajectory as
\begin{align}
r_{\mathrm{F}}(t)\equiv \mathrm{ln} \frac{\mu_{F}[n(\tau)]}{\mu_{\mathrm{B}}[\widetilde{n}(\widetilde{\tau})]},
\hspace{5mm}
r_{\mathrm{B}}(t)\equiv -r_{\mathrm{F}}(t), \label{traj-ratio}
\end{align}
where F and B stand for the forward process and the backward process, respectively.
Substituting (\ref{forwardProb}) and (\ref{backwardProb}) into (\ref{traj-ratio}), we have
\begin{align}
\mathrm{ln} \frac{\mu_{F}[n(\tau)]}{\mu_{\mathrm{B}}[\widetilde{n}(\widetilde{\tau})]}
&=\mathrm{ln} \frac{P_{0}(n_{0} )}{P_{t}(n_{N})}
- \sum_{j=1}^{N} \mathrm{ln} \frac{W_{\tau_{j}}(n_{j-1}, n_{j}; \{ \hat{\sigma}_{l}(\tau_{j}) \})}{W_{\tau_{j}}(n_{j}, n_{j-1}; \{ \hat{\sigma}_{l}(\tau_{j}) \})}\nonumber\\
&= \Delta s(t) - \Delta s_\mathrm{e}(t) = \Delta s_\mathrm{i}(t). \label{HFT1}
\end{align}
This relation immediately leads to the integrated fluctuation theorem
\begin{align}
&\sum_{n(\tau)} \mu_{\mathrm{F}}[n(\tau)] e^{-\Delta s_{\mathrm{i}}(t)} 
=:\left\langle e^{-\Delta s_{\mathrm{i}}(t)} \right\rangle_{\mathrm{F}}
=1,\label{IFT}
\end{align}
where the average $\braket{\cdots}_{\mathrm{F}}$ is the one over all forward trajectories.

The detailed fluctuation theorem also holds.
The probability that $r_{\mathrm{F}}(t)$ is equal to a value of the ``entropy production" $\Omega(t)$ is
\begin{align}
p_{\mathrm{F}}( \Omega(t) )
&=\langle \delta(\Omega(t) - \Delta s_{\mathrm{i}}(t) ) \rangle_{\mathrm{F}}\nonumber\\
&=\sum_{n(\tau)} \mu_{\mathrm{F}}[n(\tau)] \delta (\Omega(t) - r_{\mathrm{F}}(t) );
\end{align}
then (\ref{HFT1}) leads to our fluctuation theorem
\begin{align}
p_{\mathrm{F}}( \Omega(t) )
&=\sum_{n(\tau)} \mu_{\mathrm{B}}[\widetilde{n}(\widetilde{\tau})] \mathrm{e}^{ \Delta s_{\mathrm{i}}(t) } \delta (\Omega(t) - \Delta s_{\mathrm{i}}(t) )\nonumber\\
&=\mathrm{e}^{ \Omega(t) } \sum_{ \widetilde{n}(\widetilde{\tau}) } \mu_{\mathrm{B}}[\widetilde{n}(\widetilde{\tau})] \delta (\Omega(t) - r_{\mathrm{F}}(t) )\nonumber\\
&=\mathrm{e}^{ \Omega(t) } \sum_{ \widetilde{n}(\widetilde{\tau}) } \mu_{\mathrm{B}}[\widetilde{n}(\widetilde{\tau})] \delta (\Omega(t) + r_{\mathrm{B}}(t) )\nonumber\\
%&=: \mathrm{e}^{ \Omega(t) } \langle \delta(\Omega(t) + r_{\mathrm{B}}(t)) \rangle_{\mathrm{B}}\nonumber\\
&=: \mathrm{e}^{ \Omega(t) } p_{\mathrm{B}}( -\Omega(t) ),\label{DetailedFT}
\end{align}
where we used
$\sum_{\widetilde{n}(\widetilde{\tau})} = \sum_{n(\tau)}$.
Although Eq.~(\ref{DetailedFT}) is formally the same as the fluctuation theorem for the case in  a Markovian heat bath, the ``transition rate" $W_{\tau}(m, m^{\prime}; \{ \hat{\sigma}_{l}(\tau) \})$ contains the effect of non-Markovian properties in the auxiliary matrices $\{ \hat{\sigma}_{l}(\tau_{j}) \}$.

Note that the derivation of the fluctuation theorems here are totally formal; 
we need to examine whether the definition of the ``entropy production" (\ref{NMentropyproduction}) is physically legitimate.

\subsection{Points to be checked}\label{PointsToBeChecked}
It is crucial for a fluctuation theorem whether the quantities $\Delta s_\mathrm{i}(t)$ 
and $\Delta s_\mathrm{e}(t)$ 
that we defined above are indeed appropriate as the entropy production and the entropy flow, respectively.
We here list two points to be checked in Sec.~\ref{main2}.

First, as a property of the entropy production, we expect it to be zero in equilibrium. 
Therefore, $\Delta s_{\mathrm{i}}(t)$ defined in (\ref{NMentropyproduction}) should vanish there, i.e.\,
\begin{align}
\frac{P_{\tau}(n_{j}) W_{\tau}(n_{j-1}, n_{j}; \{ \hat{\sigma}_{l}(\tau) \}) }{ P_{\tau}(n_{j-1}) W_{\tau}(n_{j}, n_{j-1}; \{ \hat{\sigma}_{l}(\tau) \}) } = 1,\label{SiCondition}
\end{align}
which is nothing but a detailed balance condition.
If this condition is satisfied and the ``transition rate" $W_{\tau}(m, m^{\prime}; \{ \hat{\sigma}_{l}(\tau) \})$ is positive, then we can regard the formal equality (\ref{HFT1}) as the fluctuation theorem. 
Otherwise, (\ref{HFT1}) is an equality without any meanings of entropies. 
%Another requirement of the entropy production is that it is always non-negative in average (the average over trajectories).
As we mentioned at the end of Sec.~\ref{BDMEforHierarchy}, 
the ``transition rate" $W_{\tau}(m, m^{\prime}; \{ \hat{\sigma}_{l}(\tau) \})$ can be negative; 
we will numerically demonstrate in Sec.~\ref{main2DB} that it indeed becomes negative during a strongly non-equilibrium transient state.  In such a case, the definition (\ref{NMentropyproduction}) of the entropy production as well as (\ref{NMentropyflow}) of the entropy flow are inappropriate because 
the inside of the logarithm of (\ref{NMentropyproduction}) and (\ref{NMentropyflow}), or the left-hand side of (\ref{SiCondition}), becomes negative. 
Remarkably, we will find in Sec.~\ref{main2MR} that except for the transient case, 
the ``transition rate" $W_{\tau}(m, m^{\prime}; \{ \hat{\sigma}_{l}(\tau) \})$ is almost always positive and the ``detailed balance" is almost satisfied. 
Furthermore, we will show in Appendix \ref{markovMR} that the ``transition rate" $W_{\tau}(m, m^{\prime}; \{ \hat{\sigma}_{l}(\tau) \})$ is always positive and the equilibrium state satisfies the ``detailed balance" in the limit where the Born-Markov approximation (\ref{BornMarkovCondition}) and the rotating-wave approximation (\ref{RWA}) are applicable.

Next, following Ref.~\cite{PhysRevE.73.046129}, we define the ``heat" that flows out of the local system along a trajectory by
\begin{align}
\widetilde{q}_{\mathrm{S}}(t)
\equiv \sum_{j=1}^{N} \left( \bra{n_{j}} \hat{H}_{\mathrm{S}}(\tau_{j}) \ket{n_{j} }
- \bra{n_{j-1}} \hat{H}_{\mathrm{S}}(\tau_{j}) \ket{n_{j-1} }\right).\label{heat}
\end{align}
As a property of the entropy flow, we expect that it satisfies the following equality:
\begin{align}
\Delta s_\mathrm{e}(t) = -\beta \, \widetilde{q}_{\mathrm{S}}(t)\label{NMmicro-rev0}.
\end{align}
After defining 
$\widetilde{q}(n_{j}, n_{j-1},\tau_{j})= \bra{n_{j}} \hat{H}_{\mathrm{S}}(\tau_{j}) \ket{n_{j} }
- \bra{n_{j-1}} \hat{H}_{\mathrm{S}}(\tau_{j}) \ket{n_{j-1} }$ to be 
$\widetilde{q}_{\mathrm{S}}(t)=:\sum_{j=1}^{N} \widetilde{q}(n_{j}, n_{j-1},\tau_{j})$,
we can recast our expectation (\ref{NMmicro-rev0}) to
\begin{align}
\frac{W_{\tau_{j}}(n_{j-1}, n_{j}; \{ \hat{\sigma}_{l}(\tau_{j}) \})}
{W_{\tau_{j}}(n_{j}, n_{j-1}; \{ \hat{\sigma}_{l}(\tau_{j}) \})}
=\exp \left[ \beta \widetilde{q}(n_{j}, n_{j-1},\tau_{j}) \right].\label{NMmicro-rev}
\end{align}
The equality (\ref{NMmicro-rev}) is what we call the ``microscopic reversibility."
Note that this relation does not affect the validity of (\ref{HFT1}). 
Although the probability distributions of the initial and the final states are arbitrary in general, 
when they are both the canonical state, i.e.\,, 
$P_{0}(n_{0}) = \exp[-\beta(E(0)-F(0))]$ and $P_{t}(n_{N}) = \exp[-\beta(E(t)-F(t))]$, 
we can rewrite (\ref{HFT1}) into the form of the Crooks-type fluctuation theorem: 
\begin{align}
\frac{\mu_{F}[n(\tau)]}{\mu_{\mathrm{B}}[\widetilde{n}(\widetilde{\tau})]}
=\exp[ -\beta (\widetilde{w}(t)-\Delta F) ], \label{CrooksType}
\end{align} 
where $\Delta F \equiv F(t)-F(0)$ and we defined the ``work" 
\begin{align}
\widetilde{w}(t) \equiv \Delta E - \widetilde{q}(t), 
\end{align}
where $\Delta E \equiv E(t) - E(0)$.
Note that we can only check the consistency of the ``entropy flow" $\Delta s_\mathrm{e}(t)$ and the ``heat" $\widetilde{q}(n_{j}, n_{j-1},\tau_{j})$;
while the left-hand side of (\ref{NMmicro-rev}) is not an established quantity as the ratio of the transition rates as we mentioned at the end of Sec.~\ref{BDMEforHierarchy}, the ``heat" $\widetilde{q}(n_{j}, n_{j-1},\tau_{j})$ on the right-hand side is also a hypothetical quantity.
In Sec.~\ref{main2MR} and in Appendix \ref{markovMR}, we will show that the equality (\ref{NMmicro-rev}) holds when the Born-Markov approximation (\ref{BornMarkovCondition}) and the rotating-wave approximation (\ref{RWA}) are appropriate, although it does not hold in general.

In the following section, we will numerically examine the ``detailed balance" (\ref{SiCondition}) and the ``microscopic reversibility" (\ref{NMmicro-rev}).

\section{Numerical examination of the ``detailed balance" \\ and the ``microscopic reversibility"}\label{main2}
In order to examine the above points numerically,
let us consider the spin-boson model:
\begin{align}
&\hat{H}=\hat{H}_{\mathrm{S}}(\hat{\psi}, \hat{\psi}^{\dagger})
+\hat{H}_{\mathrm{B}}(\hat{b}_{\alpha}, \hat{b}_{\alpha}^{\dagger}) \nonumber\\
&\hspace{1cm}+\hat{H}_{\mathrm{int}}(\hat{\psi}, \hat{\psi}^{\dagger}, \hat{b}_{\alpha}, \hat{b}_{\alpha}^{\dagger})%, \\
+\hat{H}_{\mathrm{counter}}(\hat{\psi}, \hat{\psi}^{\dagger}), \label{totalSpinBoson}\\
&H_{\mathrm{S}}(\hat{\psi}, \hat{\psi}^{\dagger}) 
= \frac{\hbar \omega_{0}}{2} \hat{\sigma}_{z}
= \frac{\hbar \omega_{0}}{2} ( \hat{\psi}^{\dagger} \hat{\psi} - \hat{\psi} \hat{\psi}^{\dagger} ), \\
&\hat{H}_{\mathrm{B}}(\hat{b}_{\alpha}, \hat{b}_{\alpha}^{\dagger})
=\sum_{\alpha} \hbar \omega_{\alpha} \hat{b}_{\alpha}^{\dagger} \, \hat{b}_{\alpha}, \\
&\hat{H}_{\mathrm{int}}(\hat{\psi}, \hat{\psi}^{\dagger}, \hat{b}_{\alpha}, \hat{b}_{\alpha}^{\dagger})
= \hat{V}(\hat{\psi}, \hat{\psi}^{\dagger}) \sum_{\alpha} c_{\alpha} (\hat{b}_{\alpha} + \hat{b}_{\alpha}^{\dagger}),\\
&\hat{V}(\hat{\psi}, \hat{\psi}^{\dagger}) 
= V_{1} \hat{\sigma}_{x} + V_{2} \hat{\sigma}_{z}
= V_{1}( \hat{\psi}^{\dagger} + \hat{\psi} ) + V_{2}( \hat{\psi}^{\dagger} \hat{\psi} - \hat{\psi} \hat{\psi}^{\dagger}), \\%. \\
&\hat{H}_{\mathrm{counter}}(\hat{\psi}, \hat{\psi}^{\dagger})=\sum_{\alpha} \frac{c_{\alpha}^{2} \hat{V}(\hat{\psi}, \hat{\psi}^{\dagger})^{2}}{2m_{\alpha}\omega_{\alpha}^{2}}.
\end{align}
The operators $\hat{\sigma}_{x}$ and $\hat{\sigma}_{z}$ are the Pauli matrices and $V_{1}$ and $V_{2}$ are the corresponding coefficients.
The local system Hamiltonian $\hat{H}_{\mathrm{S}}(\hat{\psi}, \hat{\psi}^{\dagger})$ is the two-level system with its energy difference $\hbar\omega_{0}$.
We added a counter-term $\hat{H}_{\mathrm{counter}}(\hat{\psi}, \hat{\psi}^{\dagger})$ to the local system Hamiltonian in order to maintain the translation invariance of the system with respect to the heat bath \cite{citeulike:8167873,Petruccione:378945}.
We set $\hbar=1$ below.

In numerical simulation of the hierarchy equations of motion, the set of parameters $(N,K)$ in (\ref{deltaapproxK}) and (\ref{terminatorcondition}) (see Appendix \ref{HierarchyReview}) determines the accuracy of the calculation.
As we take the values of $(N,K)$ larger, the physical quantities converge to a certain value, which is numerically exact.
The present paper is based on the program \textit{nonMarkovian09}, which is distributed on the web site of Yoshitaka Tanimura \cite{tanimuraweb}.

We will mainly examine Eq.~(\ref{SiCondition}) in Sec.~\ref{main2DB} 
and examine Eq.~(\ref{NMmicro-rev}) in Sec.~\ref{main2MR}, 
while we will also investigate some other properties of the dynamics.
We again stress that the calculation is numerically exact in the sense that we did not use any approximations in deriving Eq.~(\ref{HQME1}).
In the region where the Born-Markov approximation and the rotating wave approximation are applicable, 
we can analytically investigate the behavior with the quantum optical master equation (Appendix \ref{markovMR}), 
and therefore we will compare our results to those cases.

\subsection{Thermalization after the energy measurement and the ``detailed balance"}\label{main2DB}
Let us consider the following protocol:
we measure the energy of the isolated two-level system at $\tau=0$ and find it in the ground state.
Then we connect the two-level system to the heat bath which is in the thermal equilibrium state at the inverse temperature $\beta$ and let the total system evolve without any other perturbation.
As long as the coupling between the local two-level system and the bath contains a non-vanishing term of $V_{1}$,
the local two-level system exchanges the energy with the heat bath and relaxes to a stationary state.
Since we are not perturbing the total system with any external fields, we take this stationary state as the equilibrium state.
Note that the equilibrium state of the two-level system differs from the canonical state in general;
the total system should be in the canonical state in thermal equilibrium, but the reduced system is not.

The relation that we examine here is the ``detailed balance" (\ref{SiCondition}), or equivalently
\begin{align}
\frac{W_{\tau}(a,b; \{ \hat{\sigma}_{l}(\tau) \}) }{W_{\tau}(b,a ; \{ \hat{\sigma}_{l}(\tau) \}) }
\stackrel{?}{=} \frac{P_{\tau}(a)}{P_{\tau}(b)}
\equiv \frac{\bra{a_{\tau}} \hat{\rho}(\tau) \ket{a_{\tau}} }
{ \bra{b_{\tau}} \hat{\rho}(\tau) \ket{b_{\tau}} }. \label{measDB}
%= \frac{\bra{a^{\prime}} \hat{\rho}^{\mathrm{eq}} \ket{a^{\prime}} }
%{ \bra{b^{\prime}} \hat{\rho}^{\mathrm{eq}} \ket{b^{\prime}} }, \label{measDB}
\end{align}
The set of states $\{ \ket{a_{\tau}}, \ket{b_{\tau}} \}$ is the basis that diagonalizes the reduced density matrix at time $\tau$.
We label the elements of the basis $\ket{a_{\tau}}$ and $\ket{b_{\tau}}$ in the non-decreasing order of the eigenvalues of $\hat{\rho}(\tau)$.
We denote the ratio of the ``transition rate" between the states $a_{\tau}$ and $b_{\tau}$ (the left-hand side of (\ref{measDB})) as
\begin{align}
\mathrm{DB_{L}}(\tau) := \frac{W_{\tau}(a,b; \{ \hat{\sigma}_{l}(\tau) \}) }{W_{\tau}(b,a ; \{ \hat{\sigma}_{l}(\tau) \})},  \label{measDBL}
\end{align}
while the ratio of the probability of the state at time $\tau$ (the right-hand side of (\ref{measDB})) as
\begin{align}
\mathrm{DB_{R}}(\tau) := \frac{\bra{a_{\tau}} \hat{\rho}(\tau) \ket{a_{\tau}} }
{ \bra{b_{\tau}} \hat{\rho}(\tau) \ket{b_{\tau}} }. \label{measDBR}
%\frac{\bra{a^{\prime}} \hat{\rho}^{\mathrm{eq}} \ket{a^{\prime}} }
%{ \bra{b^{\prime}} \hat{\rho}^{\mathrm{eq}} \ket{b^{\prime}} }. \label{measDBR}
\end{align}
In general, the detailed balance condition is expected to hold in the case of Markovian dynamics in thermal equilibrium.
We simulate the time evolution of the ratio $\mathrm{DB_{L}}(\tau)/\mathrm{DB_{R}}(\tau)$
in order to examine whether such a condition holds in generic cases. 
We will show that the equality holds after the system reached the equilibrium state, although the ``transition rate" (\ref{HQMErate}) can be  negative during the transient state.

\begin{figure}[th]
%\begin{minipage}{0.5\hsize}
\begin{center}
\includegraphics[width=70mm,clip]{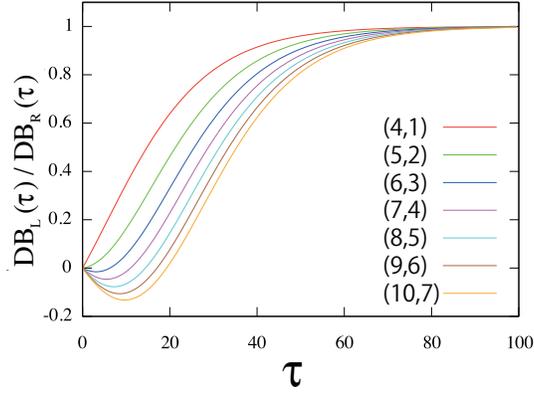}

(a)

\vspace{\baselineskip}
\end{center}
%\end{minipage}
%\begin{minipage}{0.5\hsize}
\begin{center}
\includegraphics[width=70mm,clip]{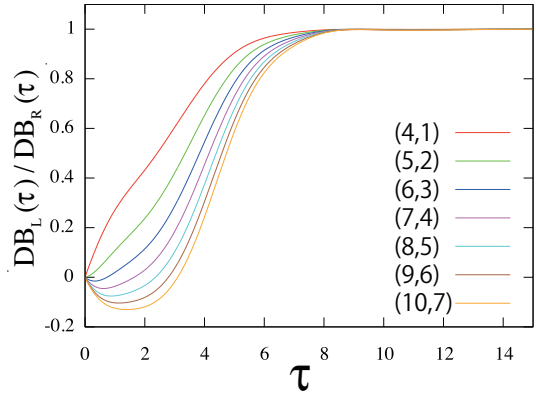}

(b)
%\label{meas6WtDBrightrho}
\end{center}
%\end{minipage}
\caption{
(color online)
The evolution of the ratio $\mathrm{DB_{L}}(\tau)/\mathrm{DB_{R}}(\tau)$ in the cases of (a) $\sigma_{x}$-coupling
and (b) $(\sigma_{x}+\sigma_{z})$-coupling after the energy measurement at $\tau=0$.
The parameter values are $\gamma=100\omega_{0}$, $\beta \omega_{0}=0.05$, and $\zeta=0.001\omega_{0}$, 
which is the case where we expect the Born-Markov approximation (\ref{BornMarkovCondition}) and the rotating-wave approximation (\ref{RWA}) to be relatively  appropriate;
$\min \left[ \gamma, 2 \pi / \beta \right] \gg \zeta$, and $\omega_{0} \gg \zeta$.
The time step is $10^{-4}$ in the simulation time.
The lines indicate, from the top to the bottom, $(N,K)=(4,1)$, $(5,2)$, $(6,3)$, $(7,4)$, $(8,5)$, $(9,6)$, and $(10,7)$.
The line converges to the case where an infinite number of auxiliary matrices are taken into account, i.e.\,the full solution.
}
\label{meas-markov}
\end{figure}

%\subsubsection{Types of the system-bath coupling}
We examine the case of $\sigma_{x}$-coupling (i.e.\,$V_{1}=1$ and $V_{2}=0$) to the bath 
as well as the case of ($\sigma_{x} + \sigma_{z})$-coupling (i.e.\,$V_{1}=V_{2}=1$).
Figures \ref{meas-markov}a and \ref{meas-markov}b show the time evolution of the ratio $\mathrm{DB_{L}}(\tau)/\mathrm{DB_{R}}(\tau)$ for the parameter region where the the Born-Markov approximation (\ref{BornMarkovCondition}) and the rotating-wave approximation (\ref{RWA}) are relatively appropriate in the cases of $\sigma_{x}$-coupling and $(\sigma_{x} + \sigma_{z})$-coupling, respectively: 
$\gamma=100 \omega_{0}$, $2\pi / \beta \simeq 120\omega_{0}$, and $\zeta=0.001\omega_{0}$.

For any cases of $\sigma_{x}$-coupling, the basis is time-independent because no off-diagonal element comes out if the state before the thermalization is diagonal.
We can understand this from (\ref{IF2}) in the Feynman-Vernon theory;
because the coupling is linear in the bath coordinate and the bath is Gaussian, the transition always occurs in even orders of $\hat{\sigma}_{x}$, and thus it causes no off-diagonal elements.
In contrast, the case of the $(\sigma_{x} + \sigma_{z})$-coupling is the one where the basis can be time-dependent.
The time evolution of $\mathrm{DB_{L}}(\tau)/\mathrm{DB_{R}}(\tau)$ in Fig.\,\ref{meas-markov}b is fluctuating compared to Fig.~\ref{meas-markov}a.
This fluctuation is due to the time evolution of the basis.

%\subsubsection{``Detailed balance" in equilibrium}
All the results of the simulations show that the ``detailed balance" (\ref{measDB}) holds once the system reaches the equilibrium state: 
$\tau \gtrsim 90$ in Fig.~\ref{meas-markov}a and $\tau \gtrsim 9$ in Fig.~\ref{meas-markov}b. 
The detailed balance is a sufficient condition, but not a necessary condition for the equilibrium state.
Therefore, this is a highly nontrivial result.
It means that the ``entropy production" that we defined in (\ref{NMentropyproduction}) satisfies the important requirement of the entropy production; it vanishes in the equilibrium state.

%\subsubsection{Positivity of the ``transition rate"}
Although everything seems fine in the equilibrium states, 
as we can see from the result of the simulation, $W_{\tau}(m,m^{\prime}; \{ \hat{\sigma}_{l}(\tau) \})$ can be negative 
(i.e.\,$\mathrm{DB_{L}}(\tau)/\mathrm{DB_{R}}(\tau)<0$) in transient states of the non-Markovian region. 
Hence we cannot always regard $W_{\tau}(m,m^{\prime})$ as the transition rate of the classical birth-death master equation.
More importantly, it implies that the ``entropy flow" (\ref{NMentropyflow}) and the ``entropy production" (\ref{NMentropyproduction}) can be ill-defined because the arguments of the logarithms becomes negative.
In contrast, we can prove that the ``transition rate" $W_{\tau}(m,m^{\prime})$ is always positive 
in the case of the quantum optical master equation (Appendix \ref{markovMR}), i.e.\,,
the Born-Markov approximation (\ref{BornMarkovCondition}) and the rotating-wave approximation (\ref{RWA}) are applicable. 
For the results of Figs.~\ref{meas-markov}a and \ref{meas-markov}b, 
we expect that (\ref{BornMarkovCondition}) and (\ref{RWA}) are appropriate as is shown above and hence the ``transition rate" would be always positive.  
The system during the transient state, however, must be in a non-Markovian region after the strong perturbation of the energy measurement.
This is indeed indicated by the fact that the ratio $\mathrm{DB_{L}}(\tau)/\mathrm{DB_{R}}(\tau)$ 
changes drastically as we include more temperature corrections of the hierarchy equations of motion, i.e.\,, increasing the number of $K$.
Except for such strongly non-equilibrium cases, our definitions of the entropy change, the entropy flow, and the entropy production seem legitimate as well as our fluctuation theorems (\ref{HFT1}), (\ref{IFT}), and (\ref{DetailedFT}).

\subsection{The ``microscopic reversibility" and other possibilities of equality}\label{main2MR}
In order to see whether the definition of the ``entropy flow" (\ref{NMentropyflow}) and the ``heat" (\ref{heat}) are legitimate, 
%and examine the possibility to rewrite the fluctuation theorem of (\ref{HFT1}) into the form of the Crooks-type or the Jarzynski equality, 
we examine the ``microscopic reversibility" (\ref{NMmicro-rev}). 
Just as the ``detailed balance," the microscopic reversibility is a general relation 
%the relation which connects the ratio of the transition rate to the heat 
in the case of Markovian dynamics.
We cannot expect such a relation to exist out of the Markovian limit; nevertheless, it is worth testing. 
We also investigate whether any quantities are related to the ratio of the ``transition rate" $\mathrm{DB_{L}}(\tau)$ other than the ``microscopic reversibility" (\ref{NMmicro-rev}).

The energy-measurement protocol of Sec.~\ref{main2DB} was suitable to see the non-equilibrium behavior, but it is difficult to investigate the non-Markovian behavior because the convergence with respect to the hierarchy of the hierarchy equations of motion gets extremely slow.

We here consider the following protocol instead: we prepare the total system in the thermal equilibrium state at time $\tau=0$ and drive the local two-level system by a sinusoidal external field for $\tau \ge 0$.
In other words, we add to the Hamiltonian of the total system (\ref{totalHamiltonian}), the Hamiltonian of the external field 
\begin{align}
&\hat{H}_{\mathrm{ext}}(\hat{\psi}, \hat{\psi}^{\dagger},\tau)
=h(\tau) \hat{\sigma}_{z}
=h(\tau)( \hat{\psi}^{\dagger}\hat{\psi} - \hat{\psi} \hat{\psi}^{\dagger})
\end{align}
with
\begin{align}
h(\tau) = A \sin(\omega_{\mathrm{ext}}\, \tau),
\end{align}
where $\omega_{\mathrm{ext}}$ is the angular frequency of the external field.
Therefore, the local system does not go to equilibrium but to a stationary state at best.
In the following, we consider the case of the $(\sigma_{x} + \sigma_{z})$-coupling to the bath.

In the present Subsection, we compare the ratio of the ``transition rates" $\mathrm{DB_{L}}(\tau)$ in (\ref{measDBL}) with the following three quantities.
First, we compare it with
 $\mathrm{DB_{R}}(\tau)$, the right-hand side of the ``detailed balance" as we defined in (\ref{measDBR}).
%\begin{align}
%\mathrm{DB_{R}}(\tau) 
%:= \frac{\bra{a_{\tau}} \hat{\rho}(\tau) \ket{a_{\tau}} }
%{ \bra{b_{\tau}} \hat{\rho}(\tau) \ket{b_{\tau}} }
%=\frac{P_{\tau}(a)}{P_{\tau}(b)}. \label{steadDBR}
%\end{align}
The equality 
\begin{align}
\mathrm{DB_{L}}(\tau)\stackrel{?}{=}\mathrm{DB_{R}}(\tau) \label{DBcheck}
\end{align}
means that the ``detailed balance" (\ref{SiCondition}) is satisfied even in the non-equilibrium state.
Note that one cannot expect the ``detailed balance" to hold out of equilibrium in general.

The second quantity that we compare with $\mathrm{DB_{L}}(\tau)$ is
\begin{align}
\mathrm{DB^{(eq)}_{R}}(h(\tau)) 
:= \frac{\bra{a^{\prime}_{\tau}} \hat{\rho}^{\mathrm{eq}}(h(\tau)) \ket{a^{\prime}_{\tau}} }
{ \bra{b^{\prime}_{\tau}} \hat{\rho}^{\mathrm{eq}}(h(\tau)) \ket{b^{\prime}_{\tau}} }, \label{steadDBeqR}
\end{align}
where $\{ \ket{a^{\prime}_{\tau}}, \ket{b^{\prime}_{\tau}} \}$ is the basis that diagonalizes the \textit{equilibrium} reduced density matrix $\hat{\rho}^{\mathrm{eq}}(h(\tau))$.
In actual calculation, we prepare the instantaneous equilibrium state $\hat{\rho}^{\mathrm{eq}}(h(\tau))$ by simulating the dynamics with $h(\tau)$ fixed.
Although $\mathrm{DB^{(eq)}_{R}}(h(\tau))$ appears to be similar to $\mathrm{DB_{R}}(\tau)$ in (\ref{measDBR}),
we calculate the ratio of the equilibrium probability with the parameter $h(\tau)$, the fixed value of the external parameter at time $\tau$.
If the density matrix of the local system is equal to the equilibrium density matrix, we have 
\begin{align}
\mathrm{DB_{R}}(\tau)\stackrel{?}{=}\mathrm{DB^{(eq)}_{R}}(h(\tau)). \label{DBeqcheck}
\end{align} 
As is shown in Sec.\,\ref{main2DB}, 
the ``detailed balance" $\mathrm{DB_{L}}(\tau)=\mathrm{DB_{R}}(\tau)$ should be satisfied in the equilibrium state, 
and thus we call that the system is in the quasi-equilibrium state when $\mathrm{DB_{L}}(\tau)\simeq\mathrm{DB^{(eq)}_{R}}(h(\tau))$ is satisfied.
We can regard that the difference between $\mathrm{DB_{L}}(\tau)$ and $\mathrm{DB^{(eq)}_{R}}(h(\tau))$ as an indicator of how close the system is to the thermal equilibrium.

The third quantity that we compare with $\mathrm{DB_{L}}(\tau)$ is $\mathrm{MR_{R}}(\tau):=\exp[-\beta \widetilde{q}(b,a,\tau)]$. 
The equality 
\begin{align}
\mathrm{DB_{L}}(\tau) \stackrel{?}{=} \mathrm{MR_{R}}(\tau) \label{MRcheck}
\end{align}
means the ``microscopic reversibility" (\ref{NMmicro-rev}).
As is shown in Appendix \ref{markovMR}, the ``microscopic reversibility" (\ref{NMmicro-rev}) holds 
in the case of the quantum optical master equation, i.e.\,, when the Born-Markov approximation (\ref{BornMarkovCondition}) and the rotating-wave approximation (\ref{RWA}) are applicable.
%If the quantum master equation is expressed as an equation of a quantum dynamical semi-group and the basis that maps the evolution of the reduced density matrix to the quantum trajectory is time independent, then the ``transition rate" $W_{\tau}(m,m^{\prime}) = \brabra{m,m} \hat{\hat{\mathcal{K}}}(\tau) \ketket{m^{\prime}, m^{\prime}}$ and the heat $\widetilde{q}(b,a,\tau)$ are completely determined by the control parameter of the Hamiltonian of the system, i.e., independent of the state itself.
%Therefore, the ``microscopic reversibility" ( (\ref{NMmicro-rev}) without the auxiliary degrees of freedom or (\ref{MarkovMicro-rev}) ) is always satisfied as long as it is satisfied at equilibrium no matter what protocol we choose. (See Appendix \ref{markovMR} for the case of the quantum optical master equation.)

Among these quantities that we compare, 
$\mathrm{DB_{L}}(\tau)$, $\mathrm{DB_{R}}(\tau)$, and $\mathrm{DB^{(eq)}_{R}}(h(\tau))$ are the quantities that are related to the birth-death master equation (\ref{HQME4}), while 
$\mathrm{MR_{R}}(\tau)=\exp[-\beta \widetilde{q}(b,a,\tau)]$ is a (hypothetical) thermodynamic quantity suggested by Esposito and Mukamel \cite{PhysRevE.73.046129}.

\begin{figure}[th]
%\begin{minipage}{0.5\hsize}
\begin{center}
\includegraphics[width=70mm,clip]{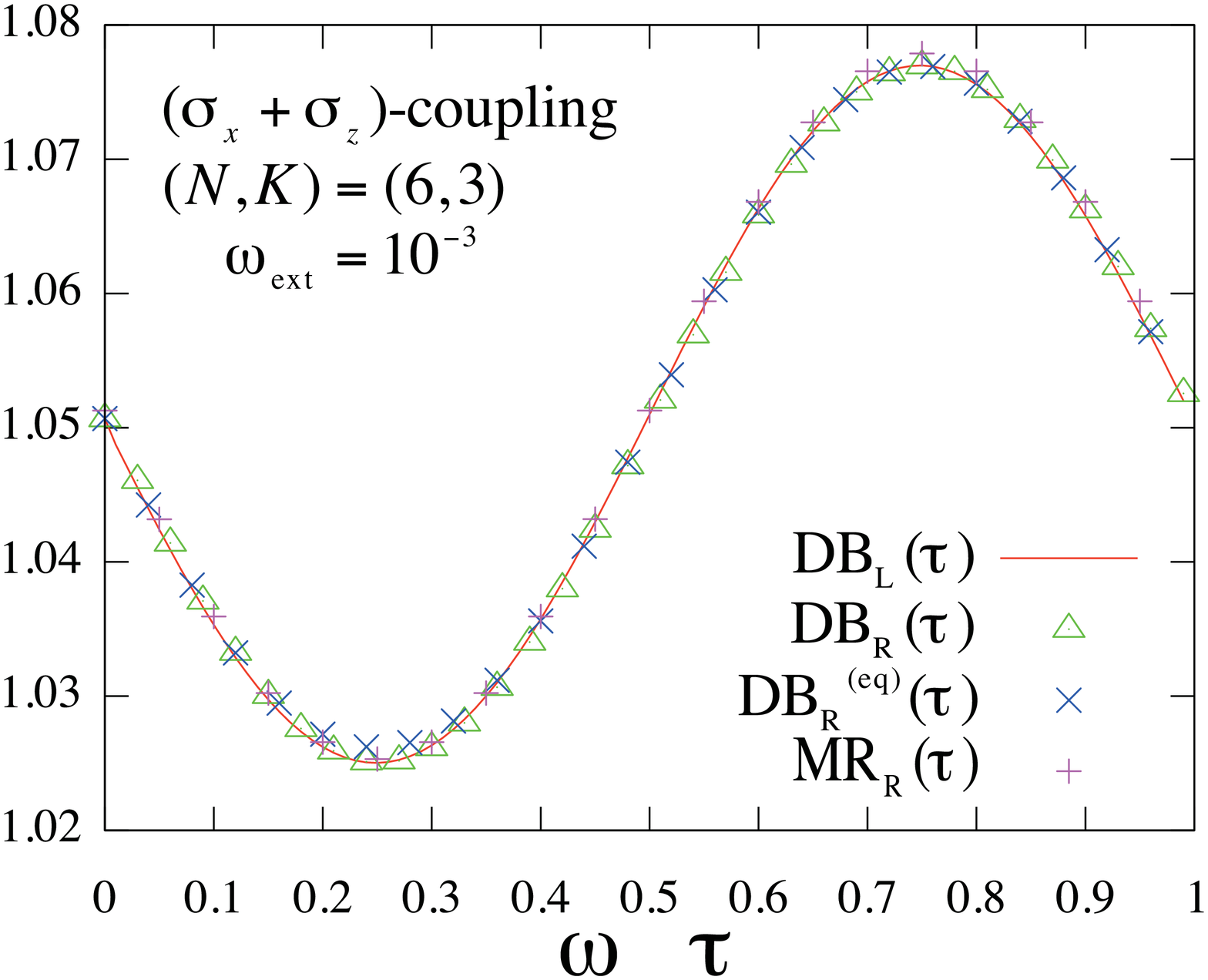}

(a)

%\label{nonop2ex10^-3MRs}
%\end{center}
%\end{minipage}
%\begin{minipage}{0.5\hsize}
%\begin{center}

\vspace{\baselineskip}
\includegraphics[width=70mm,clip]{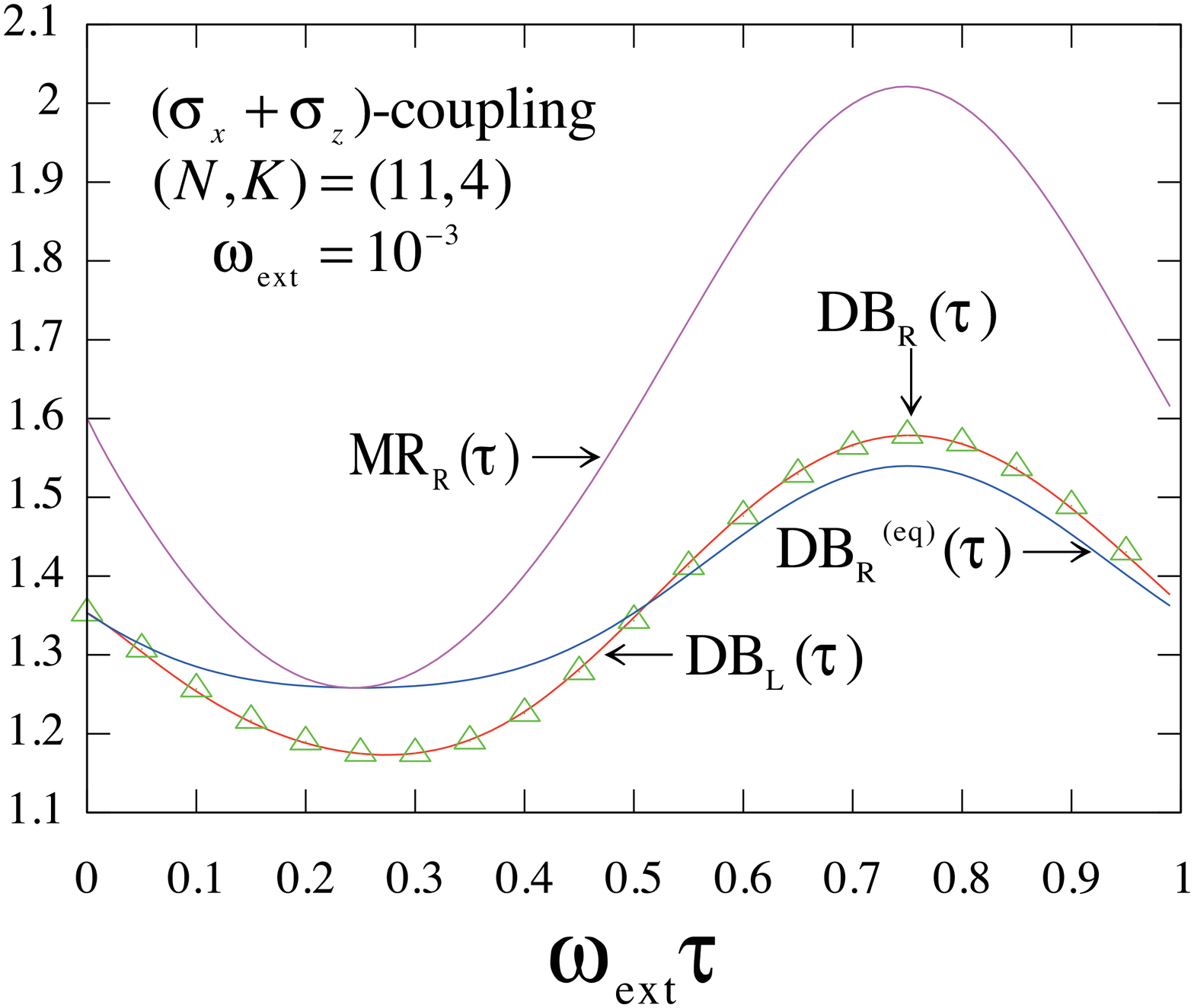}

(b)
%\label{op22ex10^-3MRs}
\end{center}
%\end{minipage}
\caption{
(color online)
(a)
The time evolution of 
$\mathrm{DB_{L}}(\tau)$ (solid line), 
$\mathrm{DB_{R}}(\tau)$ ($\triangle$ dots),
$\mathrm{DB^{(eq)}_{R}}(h(\tau))$ ($\times$ dots), and 
$\mathrm{MR_{R}}(\tau)$ (+ dots) 
for $A=0.25\omega_{0}$ and $\omega_{\mathrm{ext}}=10^{-3}$. 
The time step is $10^{-3}$ in the simulation time.
The horizontal axis is scaled as $\omega_{\mathrm{ext}} \tau$. 
The parameter values are $\gamma=100 \omega_{0}$, $\beta\omega_{0}=0.05$, and $\zeta=0.01\omega_{0}$, 
which is also the case where the Born-Markov approximation and the rotating-wave approximation are appropriate; 
$\min \left[ \gamma, 2 \pi / \beta \right] \gg \zeta$, and $\omega_{0} \gg \zeta$.
(b)
The time evolution of 
$\mathrm{DB_{L}}(\tau)$ (solid line), 
$\mathrm{DB_{R}}(\tau)$ (dots), 
$\mathrm{DB^{(eq)}_{R}}(h(\tau))$ (solid line), and 
$\mathrm{MR_{R}}(\tau)$ (solid line) 
for $A=0.25\omega_{0}$ and $\omega_{\mathrm{ext}}=10^{-3}$. 
The time step is $10^{-4}$ in the simulation time.
The horizontal axis is scaled as $\omega_{\mathrm{ext}} \tau$. 
The parameter values are $\gamma=5 \omega_{0}$, $\beta \omega_{0}=0.5$, and $\zeta=\omega_{0}$, 
which is the case where the Born-Markov approximation and the rotating-wave approximation are inappropriate. 
%; $\min \left[ \gamma, 2 \pi / \beta \right] \sim \zeta$, and $\omega_{0} = \zeta$.
In both cases, the hierarchy of $(N,K)$ is well converged.
}
\label{MRs}
\end{figure}

In order to check the consistency, 
let us first consider the parameter region where the Born-Markov approximation (\ref{BornMarkovCondition}) and the rotating-wave approximation (\ref{RWA}) are relatively appropriate: 
$\gamma=100\omega_{0}$, $2\pi/\beta\simeq 120 \omega_{0}$, and $\zeta=0.001\omega_{0}$.
The result in Fig.\,\ref{MRs}a shows that 
the state almost satisfies the ``detailed balance" (\ref{DBcheck}) as well as (\ref{DBeqcheck}),
so that the system is in a quasi-equilibrium state.
It indicates that the break of the ``microscopic reversibility"  (\ref{MRcheck}) 
is very small, and thus the ``heat" $\widetilde{q}(b,a,\tau)$ and the ``entropy flow" $\Delta s_\mathrm{e}(t)$ should  indeed satisfy the relation (\ref{NMmicro-rev0}), or equivalently, are legitimate as the heat and the entropy flow in the limit of the Born-Markov approximation (\ref{BornMarkovCondition}) and the rotating-wave approximation (\ref{RWA}). 
The result is completely consistent with the result of the quantum optical master equation in Appendix \ref{markovMR}.

Figure \ref{MRs}b is for the parameter region where the Born-Markov approximation (\ref{BornMarkovCondition}) and the rotating-wave approximation (\ref{RWA}) do not work at all:
$\gamma=5\omega_{0}$, $2\pi/\beta\simeq 12 \omega_{0}$, and $\zeta=\omega_{0}$.
Although the frequency of the external field $\omega_{\mathrm{ext}}$ is not very fast compared to the typical relaxation time of the reduced density matrix, 
Fig.\,\ref{MRs}b indicates that 
the difference of $\mathrm{DB^{(eq)}_{R}}(h(\tau))$ from $\mathrm{DB_{L}}(\tau)$ is of the same order as 
the amplitude of $\mathrm{DB^{(eq)}_{R}}(h(\tau))$ itself
and thus the system is out of equilibrium significantly.

%In Sec.~\ref{main2DB}, the ``detailed balance" (\ref{SiCondition}) was largely broken while the system is transient.
It is then remarkable that the deviation of $\mathrm{DB_{R}}(\tau)$ from $\mathrm{DB_{L}}(\tau)$ is much smaller than that of $\mathrm{DB^{(eq)}_{R}}(h(\tau))$ in the present case;
even though the system is in a strongly non-equilibrium situation, not even in the quasi-equilibrium state, the ``detailed balance" (\ref{DBcheck}) is almost satisfied. 
Note that it does not mean the ``entropy production" (\ref{NMentropyproduction}) is almost equal to zero since the system is out of equilibrium and $\dot{P}_{\tau}(n)$ in (\ref{NMentropyproduction}) is nonzero. 
The only case that we have found so far where the definition (\ref{NMentropyflow}) and (\ref{NMentropyproduction}) is inappropriate in principle is the transient state in Sec.~\ref{main2DB}.

Finally, in contrast to Fig.~\ref{MRs}a, $\mathrm{MR_{R}}(\tau)$ largely differs from $\mathrm{DB_{L}}(\tau)$, and thus the ``microscopic reversibility"  (\ref{NMmicro-rev}) does not hold at all in our sense.
It means that, in the region where the Born-Markov approximation (\ref{BornMarkovCondition}) and the rotating-wave approximation (\ref{RWA}) are inappropriate, our definitions of the ``heat" $\widetilde{q}(b,a,\tau)$ and the ``entropy flow" $\Delta s_\mathrm{e}(t)$ are not legitimate and do not satisfy the Crooks-type fluctuation theorem (\ref{CrooksType}).

\section{Conclusion}
We formally generalized the fluctuation theorems for the Markovian quantum master equation to the case of the master equation in a non-Markovian heat bath in the case of a system linearly coupled to a Gaussian heat bath with the spectrum distribution of the Drude form.
These fluctuation theorems are based on the analogy with the classical stochastic processes.
Since it was unclear whether the complete analogy holds, 
we numerically investigated the properties of the birth-death master equation (\ref{HQME4}).
The numerical results of Sec.~\ref{main2} showed that the ``detailed balance" (\ref{SiCondition}) holds in equilibrium (Sec.~\ref{main2DB}).
Therefore, the ``entropy production" (\ref{NMentropyproduction}) is indeed zero in equilibrium 
and our fluctuation theorems (\ref{HFT1}), (\ref{IFT}), and (\ref{DetailedFT}) are legitimate.
Furthermore, remarkably, the ``detailed balance" almost holds even in the weakly non-equilibrium state (Sec.~\ref{main2MR}).
We do not have a complete interpretation about this fact yet. 
The only exception that we found is that
the ``transition rate" (\ref{HQMErate}) on the quantum trajectory can be negative 
in the strongly transient dynamics after the energy measurement, 
and thus we cannot regard it as the transition rate in the sense of the classical stochastic processes. 
For such a state, the ``entropy flow" and the ``entropy production" may be ill-defined because the arguments of the logarithms of (\ref{NMentropyflow}) and (\ref{NMentropyproduction}) become negative.

From the dynamics of the sinusoidally driven system by the Zeeman field, 
we confirmed that the ``microscopic reversibility" (\ref{NMmicro-rev}) does not hold in the parameter region out of the Born-Markov limit,
although it does hold in the quantum optical master equation, which is the case where the Born-Markov approximation (\ref{BornMarkovCondition}) and the rotating-wave approximation (\ref{RWA}) are applicable (Appendix \ref{markovMR}). 
One of the possibilities that we can think of is that 
the definition of the ``heat" (\ref{heat}) is inappropriate in the non-Markovian regions.

In summary, the ``entropy production" (\ref{NMentropyproduction}) that appears in the fluctuation theorem (\ref{HFT1}) is appropriate other than in the strongly transient state. 
We cannot, however, rewrite (\ref{HFT1}) into the form of the Crooks-type fluctuation theorem with the definition of the ``heat" (\ref{heat}).
A proper definition of the heat in the non-Markovian regions might exist so that the Crooks-type fluctuation theorem may hold.

It is difficult to investigate a state very far from equilibrium numerically, because we require a large set of $(N,K)$ in order to make the result converge.
We need a more efficient methodology in order to explore the properties of the dynamics on the quantum trajectory beyond the parameter regions that we computed here.

\section*{Acknowledgements}
It is a pleasure to acknowledge discussions with M. Esposito himself.
One of us (T.K) is grateful to Y. Tanimura for fruitful comments about the hierarchy equations of motion.
The present study is supported by Grant-in-Aid for Scientific Research (B) No. 22340110 as well as by CREST, JST.

\appendix

\section{Hierarchy equations of motion}\label{HierarchyReview}
Although the hierarchy equations of motion are described well in Refs.~\cite{JPSJ.58.101,JPSJ.74.3131,JPSJ.75.082001,PhysRevA.43.4131},
we here give a brief review in order to make our description more self-contained and clarify the notations.

\subsection{The Feynman-Vernon theory}
The hierarchy equations of motion describe the dynamics of the reduced density matrix of an arbitrary local system which is linearly connected to a Gaussian heat bath, i.e.\,, the Hamiltonian of (\ref{totalHamiltonian})--(\ref{Hint}).
We assume that the initial state of the total density matrix is a product state and the density matrix of the heat bath is in the canonical state, i.e.\,,
\begin{align}
\hat{\rho}_{\mathrm{tot}}(0)&=\hat{\rho}(0) \otimes \frac{1}{Z_{\mathrm{B}}} \mathrm{e}^{-\beta \hat{H}_{\mathrm{B}}},
\hspace{5mm} Z_{\mathrm{B}}= \mathrm{Tr_{B}}\, \mathrm{e}^{-\beta \hat{H}_{\mathrm{B}}},
\end{align}
where $\mathrm{Tr_{B}}$ denotes the trace with respect to the bath degrees of freedom.
According to the Feynman-Vernon theory \cite{Feynman1963118,Grabert1988115}, the time evolution of an element of the reduced density matrix 
$\rho(\overline{\psi},\psi,\tau)
:= \bra{\overline{\psi}} \mathrm{Tr_{B}} \hat{\rho}_{\mathrm{tot}}(\tau) \ket{\psi}$
with the above initial state is expressed as
\begin{align}
&\rho(\overline{\psi}_{\mathrm{f}},\psi^{\prime}_{\mathrm{f}} ,\tau) \nonumber\\
&= \int d\overline{\psi}_{\mathrm{i}} d\psi_{\mathrm{i}} 
d\overline{\psi^{\prime}}_{\mathrm{i}} d\psi^{\prime}_{\mathrm{i}} \nonumber\\
&\hspace{5mm} \times \int^{\overline{\psi}(\tau)=\overline{\psi}_{\mathrm{f}}, \psi(\tau)=\psi_{\mathrm{f}},
\overline{\psi^{\prime}}(\tau)=\overline{\psi^{\prime}}_{\mathrm{f}}, \psi^{\prime}(\tau)=\psi^{\prime}_{\mathrm{f}} }
_{\overline{\psi}(0)=\overline{\psi}_{\mathrm{i}}, \psi(0)=\psi_{\mathrm{i}},
{\overline{\psi^{\prime}}(0)=\overline{\psi^{\prime}}_{\mathrm{i}}, \psi^{\prime}(0)=\psi^{\prime}_{\mathrm{i}} } }
 D\overline{\psi} D\psi D\overline{\psi^{\prime}} D\psi^{\prime} \nonumber\\
&\hspace{5mm} \times \mathrm{e}^{\frac{i}{\hbar}(S_{\mathrm{S}}[\overline{\psi}, \psi] - S_{\mathrm{S}}[\overline{\psi^{\prime}}, \psi^{\prime}])} \, \mathcal{F}_{\mathrm{FV}}[\overline{\psi}, \psi, \overline{\psi^{\prime}}, \psi^{\prime}]
 \rho(\overline{\psi}_{\mathrm{i}}, \psi^{\prime}_{\mathrm{i}}, 0), \label{RDM}
\end{align}
where $\overline{\psi}$ and $\psi^{\prime}$ are the coherent states,
$S_{\mathrm{S}}$ is the action of the local system, and the functional $\mathcal{F}_{\mathrm{FV}}$ is the influence functional which contains all the information of the heat bath.
Using the fact that the heat bath is Gaussian and the coupling to the local system is linear, the influence functional reads
\begin{align}
&\mathcal{F}_{\mathrm{FV}}[\overline{\psi}, \psi, \overline{\psi^{\prime}}, \psi^{\prime} ] \nonumber\\
&= \mathrm{exp} \Biggl\{
 -\frac{1}{\hbar^{2}}
 \int_{0}^{\infty}  d \omega
\int^{\tau}_{0} d\tau_{1} \, V^{\times}(\tau_{1})
\left[  -i \int^{\tau_{1}}_{0} d\tau_{2} J(\omega) \sin[\omega (\tau_{1}-\tau_{2})] V^{\circ}(\tau_{2}) \right. \nonumber\\
&\left. \hspace{6cm}+ \int^{\tau_{1}}_{0} d\tau_{2} J(\omega) \cos[ \omega(\tau_{1}-\tau_{2})] \coth\left( \frac{\beta\hbar\omega}{2} \right) V^{\times}(\tau_{2}) 
\right]
\Biggr\} \nonumber\\
&\hspace{5cm} \times
\exp \left( 
- \frac{i}{\hbar^{2}} 
\int_{0}^{\infty} d\omega \, \frac{J(\omega)}{\omega} 
\int_{0}^{\tau} d\tau_{1}\, V^{\circ}(\tau_{1}) V^{\times}(\tau_{1})
 \right), \label{IF2}
\end{align}
where
\begin{align}
& J(\omega)=\sum_{\alpha} \frac{c_{\alpha}^{2}\hbar}{2 m_{\alpha}\omega_{\alpha}}
 \delta(\omega-\omega_{\alpha}),\label{SpectrumDensity}\\
 &V^{\times}(\tau) := V(\overline{\psi}, \psi, \tau) - V(\overline{\psi^{\prime}}, \psi^{\prime}, \tau),\\
&V^{\circ}(\tau) := V(\overline{\psi}, \psi, \tau) + V(\overline{\psi^{\prime}}, \psi^{\prime}, \tau).
\end{align}

\subsection{Equations of motion}
In the case of the spectrum distribution of the Drude form (\ref{DrudeForm}),
the influence functional becomes
\begin{align}
\mathcal{F}_{\mathrm{FV}}[\overline{\psi},\psi,\overline{\psi^{\prime}},\psi^{\prime}]&=
\mathrm{exp} \left[
\int^{\tau}_{0}  d \tau_{1} \int^{\tau_{1}}_{0}  d \tau_{2} 
\Phi(\tau_{1}) \Theta(\tau_{2}) \gamma \, \mathrm{e}^{-\gamma (\tau_{1}-\tau_{2})}
\right] \nonumber\\
&\hspace{5mm}\times \prod_{k=1}^{\infty} \exp\left[ 
\int^{\tau}_{0}  d \tau_{1} \int^{\tau_{1}}_{0}  d \tau_{2} 
\Phi(\tau_{1}) \Psi_{k}(\tau_{2}) \nu_{k} \, \mathrm{e}^{-\nu_{k} (\tau_{1}-\tau_{2})}
 \right] \nonumber\\
 &\hspace{5mm} \times
\exp \left( 
- \frac{i \zeta \gamma}{2\omega_{0}} 
\int_{0}^{\tau} d\tau_{1}\, V^{\circ}(\tau_{1}) V^{\times}(\tau_{1})
 \right),
 \label{hierarchyFV1}
\end{align}
\begin{align}
&\Phi(\tau)= i V^{\times}(\tau),\label{Phi}\\
&\Theta(\tau) = \frac{i \zeta}{\beta\hbar\omega_{0}} \left[ -i \frac{\beta\hbar\gamma}{2} V^{\circ}(\tau) 
+ \frac{\beta\hbar\gamma}{2} \cot\left( \frac{\beta\hbar\gamma}{2} \right) V^{\times}(\tau) \right],\label{Theta}\\
&\Psi_{k}(\tau) = \frac{i\zeta}{\beta\hbar\omega_{0}} \frac{2 \gamma^{2}}{\nu_{k}^{2} - \gamma^{2}} V^{\times}(\tau),\label{Psi}
\end{align}
where $\nu_{k}=2\pi k/\beta\hbar$ is the Matsubara frequency.
In the form (\ref{hierarchyFV1}) of the influence functional, the time scale of each element is clear.
Hence, for $k\,(\ge K+1)$ that satisfies $\nu_{k}\gg\omega_{0}$, we may approximate
\begin{align}
&\nu_{k} \, \mathrm{e}^{-\nu_{k}(\tau - \tau^{\prime})}
\simeq \delta(\tau - \tau^{\prime})
& (k\ge K+1).  \label{deltaapproxK}
\end{align}
In other words, the Markovian approximation is valid in the levels $k\ge K+1$, but not in the lower levels.
After some calculations, under the constraint that $0<\beta\hbar\gamma/2<\pi$, we arrive at
\begin{align}
&\mathcal{F}_{\mathrm{FV}}[\overline{\psi},\psi,\overline{\psi^{\prime}},\psi^{\prime}] \nonumber\\
&\hspace{5mm}=\mathrm{exp} \Biggl\{
-\int^{\tau}_{0}  d \tau_{1} \, \left[
\Phi(\tau_{1}) \mathrm{e}^{-\gamma \tau_{1}}
\left(  -\int^{\tau_{1}}_{0}  d \tau_{2} \, \gamma \Theta(\tau_{2}) \mathrm{e}^{\gamma \tau_{2}}
\right) + \Xi(\tau_{1}) \right]
\Biggr\} \nonumber\\
&\hspace{1.5cm}\times \prod_{k=1}^{K} \exp\Biggl\{ 
-\int^{\tau}_{0}  d \tau_{1} \, \left[ \Phi(\tau_{1}) \mathrm{e}^{-\nu_{k} \tau_{1}}
\left( -\int^{\tau_{1}}_{0}  d \tau_{2} \, \nu_{k} \Psi_{k}(\tau_{2}) \mathrm{e}^{\nu_{k} \tau_{2}}
\right)
+ \Phi(\tau_{1}) \Psi_{k}(\tau_{1}) \right]
 \Biggr\}, \nonumber\\
 \label{hierarchyFV4} 
\end{align}
where we defined
%と書ける。ここで$\Xi(\tau)$を以下のように定義した。(これにはcounter termの項が入っている。)
\begin{align}
\Xi(\tau) \equiv 
\frac{\zeta}{\beta\hbar\omega_{0}}
\left[ 1- \frac{\beta\hbar\gamma}{2} \cot\left( \frac{\beta\hbar\gamma}{2} \right) \right] V^{\times}(\tau)V^{\times}(\tau)
+i \frac{\zeta}{\beta\hbar\omega_{0}} \frac{\beta\hbar\gamma}{2} V^{\circ}(\tau) V^{\times}(\tau). \label{Xi}
\end{align}

Now we introduce the following matrices:
\begin{align}
\rho^{(n)}_{j_{1}, \dots, j_{K}} 
(\overline{\psi}_{\mathrm{f}}, \psi^{\prime}_{\mathrm{f}}, \tau)
&=\int d\overline{\psi}_{\mathrm{i}} d\psi_{\mathrm{i}} 
d\overline{\psi^{\prime}}_{\mathrm{i}} d\psi^{\prime}_{\mathrm{i}}
\int^{\overline{\psi}(\tau)=\overline{\psi}_{\mathrm{f}}, \psi(\tau)=\psi_{\mathrm{f}},
\overline{\psi^{\prime}}(\tau)=\overline{\psi^{\prime}}_{\mathrm{f}}, \psi^{\prime}(\tau)=\psi^{\prime}_{\mathrm{f}} }
_{\overline{\psi}(0)=\overline{\psi}_{\mathrm{i}}, \psi(0)=\psi_{\mathrm{i}},
{\overline{\psi^{\prime}}(0)=\overline{\psi^{\prime}}_{\mathrm{i}}, \psi^{\prime}(0)=\psi^{\prime}_{\mathrm{i}} } }
 D\overline{\psi} D\psi D\overline{\psi^{\prime}} D\psi^{\prime} 
\nonumber\\
&\,\, \times \left[ \mathrm{e}^{-\gamma \tau}
\left(  -\int^{\tau}_{0}  d \tau^{\prime} \, \gamma \Theta(\tau^{\prime}) \mathrm{e}^{\gamma \tau^{\prime}}
\right) \right]^{n} \nonumber\\
&\,\,  \times \left[ \mathrm{e}^{-\nu_{k} \tau}
\left( -\int^{\tau}_{0}  d \tau^{\prime} \, \nu_{k} \Psi_{k}(\tau^{\prime}) \mathrm{e}^{\nu_{k} \tau^{\prime}}
\right) \right]^{j_{k}} \nonumber \\
& \,\, \times \mathrm{e}^{\frac{i}{\hbar} S_{\mathrm{S}}[\overline{\psi}, \psi] }
 \, \mathcal{F}_{\mathrm{FV}}[\overline{\psi},\psi, \overline{\psi^{\prime}}, \psi^{\prime}]
 \mathrm{e}^{-\frac{i}{\hbar} S_{\mathrm{S}}[\overline{\psi^{\prime}}, \psi^{\prime}] }
 \rho(\overline{\psi}_{\mathrm{i}},\psi^{\prime}_{\mathrm{i}} ,0) .
\end{align}
The element with $(n, j_{1}, \dots, j_{K})=(0, 0, \dots, 0)$ represents the original reduced density matrix.
Hereafter, we omit the subscript f for the final state.
We can express the derivative of the influence functional in terms of these matrices,
and thus the formal time derivative leads to the following hierarchy structure: 
\begin{align}
\pddt \rho^{(0)}_{0, \dots, 0} (\overline{\psi},\psi^{\prime} ; \tau)  
&= - \left( i \hat{\mathcal{L}} + \sum_{k=1}^{K} \hat{\Phi} \hat{\Psi}_{k} + \hat{\Xi} \right)
\rho^{(0)}_{0, \dots, 0} (\overline{\psi},\psi^{\prime} ; \tau)  \nonumber\\
&\hspace{5mm} - \hat{\Phi}  \rho^{(1)}_{0, \dots, 0} (\overline{\psi},\psi^{\prime} ; \tau)  
-\sum_{k=1}^{K} \hat{\Phi} \rho^{(0)}_{0, \dots, 1, \dots, 0} (\overline{\psi},\psi^{\prime} ; \tau),  
\label{hierarchyeqn0}\\
\pddt \rho^{(1)}_{0, \dots, 0} (\overline{\psi},\psi^{\prime} ; \tau)  
&= - \left( i \hat{\mathcal{L}} + \gamma + \sum_{k=1}^{K} \hat{\Phi} \hat{\Psi}_{k} + \hat{\Xi} \right)
\rho^{(1)}_{0, \dots, 0} (\overline{\psi},\psi^{\prime} ; \tau)  \nonumber\\
&\hspace{5mm} - \hat{\Phi}  \rho^{(2)}_{0, \dots, 0} (\overline{\psi},\psi^{\prime} ; \tau)  
- \gamma \hat{\Theta} \rho^{(0)}_{0, \dots, 0} (\overline{\psi},\psi^{\prime} ; \tau)  \nonumber\\
&\hspace{5mm} -\sum_{k=1}^{K} \hat{\Phi} \rho^{(1)}_{0, \dots, 1, \dots, 0} (\overline{\psi},\psi^{\prime} ; \tau),  
\label{hierarchyeqn10}\\
\pddt \rho^{(0)}_{1, 0, \dots,0} (\overline{\psi},\psi^{\prime} ; \tau)  
&= - \left( i \hat{\mathcal{L}} + \nu_{1} + \sum_{k=1}^{K} \hat{\Phi} \hat{\Psi}_{k} + \hat{\Xi} \right)
\rho^{(0)}_{1,0 \dots,0} (\overline{\psi},\psi^{\prime} ; \tau)  \nonumber\\
&\hspace{5mm} - \hat{\Phi}  \rho^{(1)}_{1,0, \dots,0} (\overline{\psi},\psi^{\prime} ; \tau)
-\sum_{k=1}^{K} \hat{\Phi} \rho^{(0)}_{1,0, \dots, 1, \dots, 0} (\overline{\psi},\psi^{\prime} ; \tau)  \nonumber\\
&\hspace{5mm} -\nu_{1} \hat{\Psi}_{1} \rho^{(0)}_{0,\dots,0} (\overline{\psi},\psi^{\prime} ; \tau),  
\label{hierarchyeqn01}\\
&\vdots \nonumber
\end{align}
\begin{align}
\pddt \rho^{(n)}_{j_{1}, \dots, j_{K}} (\overline{\psi},\psi^{\prime} ; \tau)  
&= - \left[ i \hat{\mathcal{L}} + n\gamma + \sum_{k=1}^{K} (j_{k} \nu_{k} + \hat{\Phi} \hat{\Psi}_{k} ) + \hat{\Xi} \right]
\rho^{(n)}_{j_{1}, \dots, j_{K}} (\overline{\psi},\psi^{\prime} ; \tau)  \nonumber\\
&\hspace{5mm} - \hat{\Phi}  \rho^{(n+1)}_{j_{1}, \dots, j_{K}} (\overline{\psi},\psi^{\prime} ; \tau)  
- n\gamma \hat{\Theta} \rho^{(n-1)}_{j_{1}, \dots, j_{K}} (\overline{\psi},\psi^{\prime} ; \tau)  \nonumber\\
&\hspace{5mm} -\sum_{k=1}^{K} \hat{\Phi} \rho^{(n)}_{j_{1}, \dots, j_{k}+1, \dots, j_{K}} (\overline{\psi},\psi^{\prime} ; \tau)  \nonumber\\
&\hspace{5mm} -\sum_{k=1}^{K} j_{k}\nu_{k} \hat{\Psi}_{k} \rho^{(n)}_{j_{1}, \dots, j_{k}-1, \dots, j_{K}} (\overline{\psi},\psi^{\prime} ; \tau).  
\label{hierarchyeqn}
\end{align}
In the above equations, $\hat{\Phi}, \hat{\Theta}$, $\hat{\Psi}_{k}$, and $\hat{\Xi}$ are given in (\ref{Phi}), (\ref{Theta}), (\ref{Psi}), and (\ref{Xi}) with the replacement
$V^{\times}(t) \rightarrow \hat{V}^{\times}$ and 
$V^{\circ}(t) \rightarrow \hat{V}^{\circ}$, 
where
$\hat{V}^{\times} f \equiv V f - f V$ and 
$\hat{V}^{\circ} f \equiv V f + f V$.
We can summarize (\ref{hierarchyeqn0})--(\ref{hierarchyeqn}) formally as (\ref{HQME1}).

\subsection{Terminators}
In principle, the set of above equations contains the hierarchy that continues infinitely.
Nevertheless, for the matrices with 
($n, j_{1}, j_{2}, \cdots j_{K}$) that satisfy 
\begin{align}
n \gamma + \sum_{k=1}^{K} j_{k} \nu_{k} \gg \omega_{0},\label{terminatorcondition}
\end{align}
we can approximate that such matrices obey the following equation of motion:
\begin{align}
\pddt \rho^{(n)}_{j_{1}, \dots, j_{K}} (\overline{\psi},\psi^{\prime} ; \tau)  
&\simeq - \left( i \hat{\mathcal{L}} + \sum_{k=1}^{K} \Phi \Psi_{k} + \Xi \right)
\rho^{(n)}_{j_{1}, \dots, j_{K}} (\overline{\psi},\psi^{\prime} ; \tau) .
\label{terminator2}
\end{align}
This is written solely in terms of the matrix with $(n, j_{1}, \dots, j_{K})$, and thus the hierarchy is truncated here.
The matrices which obey (\ref{terminator2}) are called the terminators.

For the numerical calculation, we set
\begin{align}
N \equiv n + \sum_{k=1}^{K} j_{k} \gg \frac{\omega_{0}}{\min(\gamma, \nu_{1})} \label{N}
\end{align}
as a number that satisfies (\ref{terminatorcondition}).

\subsection{Numerical implementation}
Analytical calculation of the hierarchy equations of motion turns out to be quite difficult in most cases,
and thus it is usual to obtain the result with numerical simulations.
In the above formulation, we can obtain the exact result
if we take $N$ in (\ref{N}) and $K$ in (\ref{deltaapproxK}) infinite so that the approximations in (\ref{terminator2}) and (\ref{deltaapproxK}) may become exact.
We cannot, however, implement this because it requires an infinite set of equations to solve.
Hence we simulate with finite values of $(N,K)$, which produces of course an approximate result.
Nevertheless, if we increase $N$ and $K$,
the physical quantity that we are calculating converges to a certain value.
We can regard this value as the numerically exact result since the simulations with the higher set of $(N,K)$ would only give negligible differences.

\section{Isolated system}\label{AppendixIsolated}
We will show in the present Appendix that the ``transition rate" $W_{\tau}(m, m^{\prime})$ vanishes for isolated systems.  
In the case of an isolated quantum system, we do not need to use the method of the hierarchy equations of motion, and thus we take over the formalism by Esposito and Mukamel \cite{PhysRevE.73.046129} (see Eqs.\,(\ref{EM1})--(\ref{EM5}) in Sec.~\ref{main1}).
The system obeys the Liouville-von Neumann equation
\begin{align}
\ddtau \ketket{\rho(\tau) }= \mathcal{\hat{\hat{L}}} \ketket{ \rho(\tau) } = -i [H(\tau), \rho(\tau)]
\end{align}
and the corresponding ``transition rate" is
\begin{align}
&W_{\tau}(m,m^{\prime}) \nonumber\\
&\equiv \langle\langle m_{\tau} | \hat{\hat{\mathcal{L} }} | m^{\prime}_{\tau} \rangle\rangle\nonumber\\
&= -i \biggl( \langle m_{\tau} | \, [
H, |m^{\prime}_{\tau} \rangle \langle m^{\prime}_{\tau} |
] \, | m_{\tau} \rangle \biggr) \nonumber\\
&= -i \biggl( \langle m_{\tau} | H |m^{\prime}_{\tau} \rangle \langle m^{\prime}_{\tau} | m_{\tau} \rangle
- \left( \langle m_{\tau} | H |m^{\prime}_{\tau} \rangle \langle m^{\prime}_{\tau} | m_{\tau} \rangle \right)^{\ast} \biggr) \nonumber\\
&=0. \label{IsoW}
\end{align}
Thus, no transitions take place.
The picture of the quantum trajectory appears obviously due to the interaction with the reservoir.

\section{``Detailed balance" and ``Microscopic reversibility" \\ of the quantum optical master equation}\label{markovMR}
In order to compare with our results in Sec.~\ref{main2}, 
we will show that the ``detailed balance" (\ref{SiCondition}), or equivalently (\ref{measDB}), and 
show that the ``microscopic reversibility" (\ref{NMmicro-rev}) holds for the spin-boson model within the Born-Markov approximation (\ref{BornMarkovCondition}) and the rotating-wave approximation.  
The condition of the rotating-wave approximation reads 
\begin{align}
\omega_{0} \gg \zeta, \label{RWA}
\end{align}
where $\omega_{0}$ is the energy spacing of the local system and $\zeta$ is the coefficient, which is related to the system-bath coupling strength $c_{\alpha}$ in Eq.~(\ref{Hint}).
These approximations are common in studies of the reservoir of the photon field, and hence we here treat such a case.
%When these approximations are applicable, we obtain a quantum master equation of the form (\ref{EM1}) (quantum optical master equation) \cite{Petruccione:378945}, 
%and therefore the discussion of this Appendix is based on the formulation given by Esposito and Mukamel \cite{PhysRevE.73.046129}.
As we did in Sec.~\ref{main2}, we discuss in the Schr\"{o}dinger picture and set $\hbar=1$.
%The interaction picture and the Schr\"{o}dinger picture coincide in the present case.

\subsection{Quantum optical master equation and the ratio of the ``transition rate"}\label{QOM}

For the Hamiltonian of the local system of $\hat{H}_{\mathrm{S}} = \omega(\tau) \sigma_{3}/2$,
the equation of motion for the reduced density matrix $\hat{\rho}(\tau)$ is given by 
the quantum optical master equation \cite{Petruccione:378945} 
\begin{align}
\ddtau \hat{\rho}(\tau) 
&= -i[\hat{H}_{\mathrm{LS}}, \hat{\rho}(\tau)] \nonumber\\
&\hspace{5mm}+ \gamma_{\mathrm{d}}(\tau)(N+1) \left( \hat{\sigma}_{-} \hat{\rho}(\tau) \hat{\sigma}_{+} 
- \frac{1}{2} \hat{\sigma}_{+} \hat{\sigma}_{-} \hat{\rho}(\tau) 
- \frac{1}{2} \hat{\rho}(\tau) \hat{\sigma}_{+} \hat{\sigma}_{-} 
\right)\nonumber\\
&\hspace{5mm}+ \gamma_{\mathrm{d}}(\tau) N 
 \left( \hat{\sigma}_{+} \hat{\rho}(\tau) \hat{\sigma}_{-} 
- \frac{1}{2} \hat{\sigma}_{-} \hat{\sigma}_{+} \hat{\rho}(\tau) 
- \frac{1}{2} \hat{\rho}(\tau) \hat{\sigma}_{-} \hat{\sigma}_{+} 
\right) \\
&=:  -i[\hat{H}_{\mathrm{LS}}, \hat{\rho}(\tau)]
+ \hathat{ \mathcal{D} } \ketket{ \hat{\rho}(\tau) }
=: ( \hathat{\mathcal{L}}_{\mathrm{LS}} + \hathat{ \mathcal{D} } ) \ketket{ \hat{\rho}(\tau) }. \label{appQOM}
\end{align}  
In deriving (\ref{appQOM}), we used the Born-Markov approximation (\ref{BornMarkovCondition}) and the rotating-wave approximation (\ref{RWA}). 
The operator $\hat{H}_{\mathrm{LS}}$ is the Hamiltonian with the Lamb-Stark shift and $\hathat{\mathcal{L}}_{\mathrm{LS}}$ is its Liouvillian,
$\hathat{ \mathcal{D} }$ is referred to as the dissipator,
$\hat{\sigma}_{+}=(\hat{\sigma}_{x} + i \hat{\sigma}_{y})/2$, $\hat{\sigma}_{-}=(\hat{\sigma}_{x} - i \hat{\sigma}_{y})/2$, 
%\begin{align}
%&\hat{\sigma}_{+}=
%\begin{pmatrix}
%0&1\\
%0&0
%\end{pmatrix},
%&\hat{\sigma}_{-}=
%\begin{pmatrix}
%0&0\\
%1&0
%\end{pmatrix},
%\end{align}
and
\begin{align}
&\gamma_{\mathrm{d}}(\tau)= \frac{4 \omega (\tau)^{3} |\vec{d}|^{2} }{3 c^{3}},  &N:=N(\omega(\tau)) = \frac{1}{\exp\bigl(\beta\omega(\tau)\bigr)-1}.
\end{align}
The factor $c$ is the speed of light and
$\vec{d}$ is the off-diagonal element of the dipole operator,
\begin{align}
\vec{D}_{\mathrm{dipole}}(\tau) = \vec{d} \hat{\sigma}_{-} \mathrm{e}^{-i\omega(\tau)\tau} 
+ \vec{d}^{\ast} \hat{\sigma}_{+} \mathrm{e}^{i\omega(\tau) \tau}.
\end{align}
This corresponds to the case of $\sigma_{x}$-coupling in Sec.~\ref{main2}.
In the current model and approximations, the $(\sigma_{x}+\sigma_{z})$-coupling would only cause the energy shift from the case of $\sigma_{x}$-coupling.

Since Eq.~(\ref{appQOM}) is of the form (\ref{EM1}), we follow Esposito and Mukamel \cite{PhysRevE.73.046129} here.
The general solution of the density matrix is expressed as
\begin{align}
\hat{\rho}(\tau) &= 
 \mathrm{e}^{-i \tau \hat{H}_{\mathrm{LS}} }
\frac{1}{2} ( 1+ \braket{\vec{\sigma}(\tau)} \cdot \vec{\sigma} )
 \mathrm{e}^{i \tau \hat{H}_{\mathrm{LS}} } \nonumber\\
&= \mathrm{e}^{-i \tau \hat{H}_{\mathrm{LS}} }
\begin{pmatrix}
\frac{1}{2} (1+\braket{\sigma_{3}(\tau)})&\braket{\hat{\sigma}_{-}(\tau)}\\
\braket{\hat{\sigma}_{+}(\tau)}&\frac{1}{2} (1-\braket{\sigma_{3}(\tau)})
\end{pmatrix}
 \mathrm{e}^{i \tau \hat{H}_{\mathrm{LS}} }. \label{rho-evo1}
\end{align}
The angular bracket means $\braket{\cdots}=\mathrm{Tr_{\mathrm{S}}}( \cdots \hat{\rho}(\tau) )$, where 
$\mathrm{Tr_{\mathrm{S}}}$ is the trace over the local system. 
Each of these averaged values evolves as 
\begin{align}
&\ddtau \braket{\sigma_{1}(\tau)} = -\frac{\gamma_{\mathrm{d}}(\tau)[2N+1]}{2} \braket{\sigma_{1}(\tau)},\\
&\ddtau \braket{\sigma_{2}(\tau)} = -\frac{\gamma_{\mathrm{d}}(\tau)[2N+1]}{2} \braket{\sigma_{2}(\tau)},\\
&\ddtau \braket{\sigma_{3}(\tau)} = - \gamma_{\mathrm{d}}(\tau)[2N+1] \braket{\sigma_{3}(\tau)} - \gamma_{\mathrm{d}}(\tau).\label{rho-evo2}
\end{align}
Since the coupling is weak under the present approximations, the initial equilibrium state is well approximated as
\begin{align}
\hat{\rho}(0) = 
\begin{pmatrix}
\mathrm{e}^{-\frac{1}{2}\beta \omega(0)}&0\\
0&\mathrm{e}^{\frac{1}{2}\beta \omega(0)}
\end{pmatrix}.
\end{align}
Thus, the density matrix in the interaction picture is a diagonal matrix all the time,
which means that the time-dependent basis (\ref{td-basis})
% $\{\ket{a_{t}}, \ket{b_{t}} \}$ 
becomes time independent:
\begin{align}
&\ket{a_{\tau}}=
\begin{pmatrix}
a_{1}\\
a_{2}
\end{pmatrix}
=\begin{pmatrix}
1\\
0
\end{pmatrix},
&\ket{b_{\tau}}=
\begin{pmatrix}
b_{1}\\
b_{2}
\end{pmatrix}
=\begin{pmatrix}
0\\
1
\end{pmatrix}.
\end{align}

Because $\brabra{b_{\tau}}\hathat{\mathcal{L}}_{\mathrm{LS}}\ketket{a_{\tau}}=0$
as we see from (\ref{IsoW}), 
the ratio of the ``transition rate" $W_{\tau}(b_{\tau},a_{\tau})$ in (\ref{EM5}) reads
\begin{align}
W_{\tau}(b_{\tau},a_{\tau}) &\equiv \brabra{ b_{\tau} } \hathat{\mathcal{D}} \ketket{ a_{\tau} }\\
&=\gamma_{\mathrm{d}} N \bigl( |a_{1}|^{2} |b_{2}|^{2} + |a_{2}|^{2} |b_{1}|^{2} \bigr)
+ \gamma_{\mathrm{d}} |a_{2}|^{2} |b_{1}|^{2}. \label{MarkovTransitionRate}
\end{align}
The second term is asymmetric with respect to the process reversal $a_{\tau} \leftrightarrow b_{\tau}$. 
We also see that the ``transition rate" $W_{\tau}(b_{\tau},a_{\tau})$ is real and positive unlike the general expression (\ref{HQMErate}). 
Substituting the specific form $N=(e^{\beta \omega(\tau)} - 1)^{-1}$ into (\ref{MarkovTransitionRate}), 
we have for Eq.~(\ref{measDBL})
\begin{align}
\mathrm{DB_{L}}(\tau)
&=\frac{W_{\tau}(a,b)}{W_{\tau}(b,a)} \nonumber\\
&= \frac{
e^{\beta \omega(\tau)} |a_{2}|^{2} |b_{1}|^{2} + |a_{1}|^{2} |b_{2}|^{2}
}{
e^{\beta \omega(\tau)} |a_{1}|^{2} |b_{2}|^{2} + |a_{2}|^{2} |b_{1}|^{2}
}
=\mathrm{e}^{-\beta \omega (\tau)}. \label{MarkovMicro-rev}
\end{align}

\subsection{``Detailed balance"}\label{AppendixMarkovDB}
Let us consider the time evolution from the ground state of the two-level system with $\omega(\tau)=\omega_{0}$, just as in Sec.~\ref{main2DB}.
According to (\ref{rho-evo1})--(\ref{rho-evo2}), the time evolution of the density matrix $\hat{\rho}(\tau)$ reads
\begin{align}
\hat{\rho}(\tau) = 
\begin{pmatrix}
\frac{1}{2} (1+\braket{\sigma_{3}(\tau)})&0\\
0 & \frac{1}{2} (1-\braket{\sigma_{3}(\tau)})
\end{pmatrix}
\end{align}
with
\begin{align}
\braket{\sigma_{3}(\tau)} =
\frac{2(N(\omega_{0})+1)}{2N(\omega_{0})+1} \mathrm{e}^{-\gamma_{\mathrm{d}}(2N(\omega_{0})+1)\tau} 
-\frac{1}{2N(\omega_{0})+1}.
%\mathrm{e}^{-\gamma_{\mathrm{d}}(2N+1)\tau} 
%- \gamma_{\mathrm{d}} \int^{\tau}_{0} ds\, \mathrm{e}^{-\gamma_{\mathrm{d}}(2N+1)(\tau-s)}.
\end{align}
Thus we have for Eq.~(\ref{measDBL})
\begin{align}
\mathrm{DB_{R}}(\tau) = \frac{\bra{a_{\tau}} \hat{\rho}(\tau) \ket{a_{\tau}} }
{ \bra{b_{\tau}} \hat{\rho}(\tau) \ket{b_{\tau}} }
=\frac{1+\braket{\sigma_{3}(\tau)}}{1-\braket{\sigma_{3}(\tau)}}. 
\end{align}
As $\tau\rightarrow\infty$, we have $\mathrm{DB_{R}}(\tau) \rightarrow \exp(-\beta \omega_{0})$. 
Therefore, the ``detailed balance" (\ref{SiCondition}), or (\ref{measDB}) holds in equilibrium under the present approximations.

\subsection{``Microscopic reversibility"}\label{AppendixMarkovMR}
For the case where the local system is driven by the Zeeman magnetic field as in Sec.~\ref{main2MR},
the right-hand side of (\ref{NMmicro-rev}) reads
\begin{align}
&\bra{b_{\tau}} H_{\mathrm{S}}(\tau) \ket{b_{\tau}} = -\frac{\hbar\omega(\tau)}{2}, \hspace{5mm} 
\bra{a_{\tau}} H_{\mathrm{S}}(\tau) \ket{a_{\tau}}=\frac{\hbar\omega (\tau)}{2},\\
&\exp (\beta[ \bra{b_{\tau}} H_{\mathrm{S}}(\tau) \ket{b_{\tau}} - \bra{a_{\tau}} H_{\mathrm{S}}(\tau) \ket{a_{\tau}} ]) = \mathrm{e}^{-\beta \hbar \omega (\tau)}. \label{QOMDBR}
\end{align}
Comparing (\ref{QOMDBR}) with (\ref{MarkovMicro-rev}), we see that the ``microscopic reversibility" (\ref{NMmicro-rev}) holds in this case.
Note, however, that it does not generally hold even if we use the Born-Markov approximation and the rotating-wave approximation, when we drive the system with the external field other than the Zeeman magnetic field.

%For the case of the external field which alters the diagonal elements of the reduced system, e.g. the Zeeman magnetic field on a spin:
%\begin{align}
%\hat{H}_{\mathrm{S}}(\tau) = 
%\begin{pmatrix}
%\hbar\omega(\tau)&0\\
%0&-\hbar \omega(\tau)
%\end{pmatrix},
%\hspace{5mm}
%\hbar \omega(\tau) = \frac{\hbar \omega_{0}}{2}+h(\tau).
%\end{align}

\bibliographystyle{apsrev}
\bibliography{QMEthermoBib}

\end{document}